\documentclass[aps,showpacs,12pt,amssymb,preprintnumbers,amsmath,amsfonts]{revtex4}
\usepackage{latexsym}
\usepackage{graphicx}

\usepackage{color}
\usepackage{amsbsy}
\usepackage{amstext}
\usepackage{amssymb}
\usepackage{cancel}
\newcommand{\be}{\begin{equation}}
\newcommand{\ee}{\end{equation}}
\newcommand{\ba}{\begin{eqnarray}}
\newcommand{\ea}{\end{eqnarray}}
\newcommand{\nn}{\nonumber}
\usepackage{slashed}

\begin{document}


\title{The on-shell effective field theory: a systematic tool to compute power corrections to the
hard thermal loops}

\author{Cristina Manuel}
\email{cmanuel@ice.cat}
\affiliation{Instituto de Ciencias del Espacio (IEEC/CSIC) C. Can Magrans s.n., 08193 Cerdanyola del Vall\`es, Catalonia, Spain}
\author{Joan Soto}
\email{joan.soto@ub.edu}
\affiliation{Departament d'Estructura i Constituents de la Mat\`eria 
                   and Institut de Ci\`encies del Cosmos,
        Universitat de Barcelona,
        Mart\'\i $\,$ i Franqu\`es 1, 08028 Barcelona, Catalonia, Spain.}
\author{Stephan Stetina}
\email{stetina@uw.edu}
\affiliation{Institute for Nuclear Theory, University of Washington, Seattle, WA 98195, USA}

\begin{abstract}

We show that effective field theory techniques can be efficiently used to compute power corrections  to the hard thermal loops (HTL) 
 in a high temperature T expansion. To this aim, we use the recently proposed on-shell effective field theory (OSEFT), which describes the quantum fluctuations around on-shell degrees of freedom.
 We provide the OSEFT Lagrangian up to third order in the 
energy expansion for QED, and use it for the computation of power corrections to the retarded photon polarization tensor for soft external momenta.
Here soft denotes a scale of order $eT$, where $e$ is the gauge coupling constant. We develop the necessary techniques to perform these computations, and study the contributions to the polarization tensor proportional to $e^2 T^2$, $e^2 T$ and $e^2 T^0$. The first one describes the HTL contribution, the second one vanishes, while the third one provides corrections of order $e^2$ to the soft photon propagation. 
We check that the results agree with the direct calculation from QED, up to local pieces, as expected in an effective field theory.

\end{abstract}

\maketitle

\section{Introduction}

The physics of  QED and QCD plasmas at high temperature $T$ is extremely rich \cite{Blaizot:2001nr}.  In the early 90's it was discovered that the soft energy and momentum scales of these plasmas, where soft denotes a scale of order $eT$, and $e$ is the gauge coupling constant, are properly described by the so called 
 Hard Thermal Loop (HTL) effective field theory. HTLs were first found out by  extracting from one-loop Feynman diagrams their leading behavior for soft external momenta \cite{Pisarski:1988vd,Braaten:1989mz,Frenkel:1989br},
 which arises from the contribution of the so called hard scales (of order $T$) circulating in the loop.
 For  soft scales,  they are as relevant as the bare propagators or vertices of the theory, and  HTLs have to be resummed.
Although different derivations  of the HTL diagrams were given,
it was soon realized that they could be understood in terms of the classical propagation of the on-shell particles of the QED or QCD plasmas
\cite{Silin:1960pya,Kelly:1994ig,Kelly:1994dh}. 
The HTL effective field theory has been used for a large variety of computations of both static and dynamical properties of thermal plasmas (see for example
 Ref.\cite{Haque:2014rua}),
while  the static properties in the high T limit of QED and QCD have been typically studied 
with the use of dimensional reduced effective field theories \cite{Appelquist:1981vg,Kajantie:1995dw,Braaten:1995cm}.
One of the aims of this paper is  to show that  effective field theory techniques can also be used to compute power corrections to the HTLs,
 as arising from the hard scales in the plasma.

 The concept of effective field theory  (EFT) is widely and successfully  used in Physics. It relies on the idea that in order to discuss relevant phenomena at a given energy scale, it is enough to identify the degrees of freedom that operate at that scale, and uncover the Lagrangian that governs their dynamics. The Lagrangian is organized in operators of increasing dimension over powers of the high energy scales, so that all the information on the high energy scales (beyond the explicit powers) is encoded in the matching coefficients of these
operators. The matching coefficients are obtained by enforcing the EFT to be equivalent to the fundamental theory at a given order
in the ratio of scales and/or in some small parameter, typically a coupling constant. 
Nowadays, a large number of EFTs have been derived at zero temperature, from which we will only quote the ones that have been relevant for the present work. 
 High Density Effective Theory (HDET)
describes the quantum fluctuations around the Fermi level of a finite density system, being the chemical potential the high energy scale \cite{Hong:1998tn}.
In Non-Relativistic QED/QCD (NRQED/NRQCD) \cite{Caswell:1985ui} the  high energy scale is the mass of the heavy particles and the low energy scales the remaining ones in a non-relativistic bound state. Heavy Quark Effective Theory (HQET) \cite{Eichten:1989zv,Georgi:1990um,Grinstein:1990mj} may be considered the simplest particular case of NRQCD, in which the only low energy scale is $\Lambda_{QCD}$, the typical hadronic scale. The construction of HQET is formally very similar to the one of HDET, and it was a source of inspiration
 for the so called Large Energy Effective Theory (LEET) \cite{Dugan:1990de}. Soft-Collinear Effective theory (SCET) \cite{Bauer:2000ew,Bauer:2000yr} may be regarded as a completion of the latter, in which the high energy scale is a dynamical variable and hence the matching coefficients are dynamical functions rather than functions of fixed parameters. This feature will be shared by the EFT used in this work. It first appeared in Potential NRQCD/NRQED (pNRQED/pNRQCD) \cite{Pineda:1997bj}, 
in which the quantum mechanical potentials are regarded as position-dependent matching coefficients. In recent years, some of the EFTs above have been combined with
the thermal EFTs in order to study the thermal properties of non-relativistic bound states \cite{Escobedo:2008sy,Brambilla:2008cx,Brambilla:2010vq,Escobedo:2010tu,Escobedo:2011ie,Escobedo:2013tca} and jets \cite{Benzke:2012sz}.

In this manuscript we will show that the recently proposed on-shell effective field theory (OSEFT), see Ref.~\cite{Manuel:2014dza}, is a systematic and powerful tool
to extract power corrections to the HTLs. As an example we focus here in studying the retarded polarization tensor of QED. The OSEFT was 
used in Ref.~\cite{Manuel:2014dza}
to provide a derivation of  chiral kinetic theory at finite temperature \cite{Son:2012wh,Stephanov:2012ki,Son:2012zy,Manuel:2013zaa}. 
However, its possible applications have a much wider scope.
The OSEFT is meant to describe physical phenomena dominated by almost on-shell particles. Here, for simplicity, we will restrict
ourselves to the case of QED with massless fermions.
The formalism can be generalized
to deal with on-shell massive particles or with non-Abelian interactions. Starting from the QED Lagrangian we derive the Lagrangian describing the (small) quantum fluctuations around the on-shell degrees of freedom, which can be expanded as a series in $1/p$, where $p$ is the energy associated to the on-shell degrees of freedom in a given frame. In a thermal plasma, for
$p \sim T$ in the rest frame of the plasma,  our formalism not only allows us to easily extract the HTLs, but also  
 corrections to them expanded in powers of $1/T$.
In this paper we develop the techniques to perform these computations, and extract the first corrections to the HTL associated with  the retarded photon polarization tensor. We also check
explicitly that the results obtained from the OSEFT agree with those obtained with full QED, at the order of accuracy we work, up to local counterterms, as it should be the case in an
EFT.

This paper is structured as follows.  In Sec.~\ref{sec:OSEFT} we present the rationale behind the OSEFT, and how to derive the effective Lagrangian associated with
the quantum fluctuations around on-shell particles and antiparticles. In Sec.~\ref{sec:Lag} we give the explicit form of the effective Lagrangians up to order $1/p^3$ in
the energy expansion, after performing local field redefinitions, which facilitate the computations carried out in this manuscript. In Sec.~\ref{sec:propagators} 
we present the propagators for the particle and antiparticle quantum fluctuations in a thermal bath, in the so called real time formalism. Sec.~\ref{sec:polarization} is devoted
to  the computation of the retarded photon polarzation tensor with the OSEFT. After introducing the two relevant topologies - bubble and tadpole diagrams- and making some generic comments on how to organize the calculation, we show results at
order $e^2 T^2$, $e^2 T$, and $e^2 T^0$, in Secs.~\ref{Orderp1}, \ref{Orderp2} and \ref{Orderp3}, respectively. 
In Sec.~\ref{Orderp1} we obtain the standard HTL result and in Sec.~\ref{Orderp2} we show that there is no contribution at order $e^2 T$. In Sec.~\ref{Orderp3} we present the contribution of the bubble and tadpole diagrams separately,  and pin point the inherent ambiguities of the latter at this order. We close with a discussion in Sec.~\ref{discussion}. Appendices A and C contain technical details. Appendix B shows how the calculations can be performed in the imaginary time formalism, and in Appendix D we carry out the calculation directly from QED in order to check the reliability of the OSEFT.

We use natural units $\hbar= c = k_B =1$, metric conventions $g^{\mu \nu} = (1, -1,-1,-1)$, and boldface letters to denote 3-dimensional vectors.

\section{The on-shell effective field theory}
\label{sec:OSEFT}


In this section we review how to construct the basic effective action of the OSEFT
 first introduced in Ref.~\cite{Manuel:2014dza}.
For the computation of different 
physical observables dominated by the contribution of almost on-shell fermions it is convenient to construct an 
EFT
 where the role of the quantum fluctuations is clearly singled out. Let us recall that  
the propagation of an on-shell massless fermion is  described by its energy $E= p$, with $p > 0$, and the four light-like velocity $v^\mu= (1, {\bf v})$, where
${\bf v}$ is three-dimensional unit vector.  
Hence, for a fermion close to be on-shell, its four momentum can be expressed as 
\be
q^\mu = p v^\mu + k^\mu \ ,\label{eq:Qpart}
\ee
where $k^\mu$  is the residual momentum ($k^\mu \ll p$), {\it i.e}. the part of the momentum which makes $q^\mu$ slightly off-shell. 
 
A similar decomposition of the momentum for almost on-shell antifermions can be done  as follows
\begin{equation}
q^{\mu}=-p\tilde{v}^{\mu}+k^{\mu}\,\label{eq:QAntiPart}
\end{equation}
where  $\tilde{v}^{\mu}=(1,-\bf{v})$~. 

We will  apply these splittings when writing the Lagrangian of almost on-shell fermions, as then
\begin{equation}
\mathcal{L}=\sum_{p, {\bf v}}\mathcal{L}_{p, {\bf v}}\,,\qquad \mathcal{L}_{p,{\bf v}}=\bar{\psi}_{\bf v}\gamma\cdot iD\psi_{\bf v}\,,\qquad iD_\mu=i\partial_\mu+eA_\mu \,. \label{eq:Lv}
\end{equation}

The electromagnetic field above is assumed to contain soft momenta only ($l_\mu \ll p$). The precise meaning of the sum shown in Eq.~(\ref{eq:Lv})
 will be given later on.

The Dirac field in Eq.~(\ref{eq:Lv}) can be written factoring out its $p$-dependence
\begin{equation}
\psi_{\bf v}=e^{-ipv\cdot x}\left(P_{ v}\chi_{v}(x)+P_{\tilde v}H_{\tilde v}^{(1)}(x)\right)+e^{ip\tilde{v}\cdot x}\left(P_{\tilde v}\xi_{\tilde v}(x)+P_v H_{v}^{(2)}(x)\right) \ ,
\label{eq:Fields}
\end{equation}
where 
%
\begin{eqnarray}
P_v & = & \frac{1}{2}\gamma\cdot v\,\gamma_{0}\,,\label{eq:PProj}\\
P_{\tilde v} & = & \frac{1}{2}\gamma\cdot\tilde{v}\,\gamma_{0}.\label{eq:AProj}
\end{eqnarray}
are the particle and antiparticle projectors, respectively. The fields
$\chi_{v}(x)$ and $\xi_{\tilde v}(x)$ contain soft momenta only ($k_\mu \ll p$), whereas $H_{v}^{(1)}(x)$ and $H_{v}^{(2)}(x)$ contain generic off-shell  momenta.
Then after integrating out the $H_{\tilde v}^{(1)}$ and $H_{v}^{(2)}$ fields
 (see Ref. \cite{Manuel:2014dza} for details)
one obtains the following effective Lagrangian 
\begin{eqnarray}
\mathcal{L}_{p,{\bf v}}\ & = & \mathcal{L}_{p, v} + \widetilde {\mathcal{L}}_{p,{\tilde v}} 
\nonumber \\
& = &
\chi_{v}^{\dagger}(x)\left(i\, v\cdot D\,
+i \slashed{D}_{\perp} \frac{1}{2 p + i \tilde{v}\cdot D   }\,i \slashed{D}_{\perp}
\right)\chi_{v}(x) \nonumber \\
&+& \xi_{\tilde v}^{\dagger}(x)\left(i\, \tilde{v}\cdot D\,
+i \slashed{D}_{\perp} \frac{1}{ - 2 p + i  v\cdot D   }\,i \slashed{D}_{\perp} \right)\xi_{\tilde v}(x)
\,.\label{eq:Leff}
\end{eqnarray}
where  $\slashed{D}_{\perp} = P^{\mu \nu}_{\perp} \gamma_\mu D_\nu$,
and
\be
P^{\mu \nu}_{\perp} = g^{\mu \nu} - \frac 12 \left( v^\mu {\tilde v}^\nu +v^\nu {\tilde v}^\mu\right) \ ,
\ee
is minus the transverse projector to ${\bf v}$, written in covariant form. Note that $D^0_\perp=0$ and, in our conventions, $k^2_\perp = P^{\mu \nu}_{\perp} k_\mu k_\nu = - {\bf k}^2_\perp$.

In the OSEFT  the particle and antiparticle degrees of freedom, described by the $\chi$ and $\xi$ fields, respectively, are totally decoupled.
That is why the EFT techniques  employed here can be seen as the quantum field theory
counterpart of the Foldy-Wouthuysen diagonalization methods employed at the level of the first quantized Dirac Hamiltonian \cite{Foldy:1949wa}.

Note also that the antiparticle part
of the Lagrangian keeps the same structure as the particle part, as the two are equivalent if one performs  the changes
$p \leftrightarrow - p$ and $v^\mu \leftrightarrow {\tilde v}^\mu$ (or ${\bf v} \leftrightarrow -{\bf v}$) . This is a reflection of the CP symmetry of the underlying theory.

\subsection{Effective Lagrangian up to third power in the energy expansion}
\label{sec:Lag}


The effective theory just presented allows us to assess the effect of the quantum fluctuations  to different
processes dominated by almost on-shell fermions in an expansion in powers of $1/p$.  In order to do so
one simply has to expand in $1/p$ the Lagrangian Eq.~(\ref{eq:Leff}). The first two terms in this expansion were explicitly considered in Ref.~\cite{Manuel:2014dza}. They read

 \ba
 \label{Lan-0}
 \mathcal{L}^{(0)}_{p, v}\ &= & \chi_{v}^{\dagger} \left(i\, v\cdot D\,\right)\chi_{v} \ , \\
 \label{Lan-1}
  \mathcal{L}^{(1)}_{p, v}\ &= & - \frac {1}{2 p}\chi_{v}^{\dagger} \, (\slashed{D}_{\perp})^{2} \, \chi_{v} = 
   - \frac {1}{2 p}\chi_{v}^{\dagger} \,\left( D_{\perp}^{2} - \frac e2 \sigma^{\mu \nu}_\perp F_{\mu \nu} \right)\chi_{v} \ ,
 \ea
 where $ \sigma^{\mu \nu}_\perp = P^\mu_{\perp \alpha} P^{\nu}_{\perp \beta} \sigma^{\alpha \beta}$, and $\sigma^{\mu \nu} = \frac i2 [\gamma^\mu,\gamma^\nu]$. We will focus on the Lagrangian for the particle fluctuations, the Lagrangian for the antiparticle fluctuations is easily obtained after performing
the changes $p \leftrightarrow - p$ and $v^\mu \leftrightarrow {\tilde v}^\mu$.
 
 The interaction terms generated at order
 $1/p^{2}$, and higher, contain temporal derivatives. In order to simplify the computations
 of different Feynman diagrams at this and higher orders it is convenient to perform  local
 field redefinitions, such that only  the leading order Lagrangian contains
   temporal derivatives acting on the fermionic fields. This is a standard procedure in non-relativistic effective theories \cite{Manohar:1997qy}.
 Thus,  if at order $1/p^2$ we make the  field redefinition
\begin{equation}
\chi_{v}\rightarrow\chi_{v}^{\prime}=\left(1 +\frac{   \slashed{D}_{\perp}^{2}}{8p^{2}}\right)\chi_{v}\ ,
\end{equation}
the Lagrangian at this order reads
%
%
\begin{equation}
\label{Lan-2}
\mathcal{L}^{(2)}_{p, v}\ = \frac{1}{8p^{2}} \chi_{v}^{\prime\dagger} \Big(\left[ \slashed{D}_{\perp}\,,\,\left[i\tilde{v}\cdot D\,,\,\slashed{D}_{\perp}\right]\right] -
\left\{ ( \slashed{D}_{\perp})^{2},\,\left(iv\cdot D-i\tilde{v}\cdot D\right)\right\}  \Big) \chi_{v}^{\prime} \ ,
\end{equation}
 where $\left\{ \,,\,\right\} $ denotes the anti-commutator.
This Lagrangian  is similar, though not identical, to the Lagrangian obtained
for massive fermions in a non relativistic $1/m$ expansion \cite{Manohar:1997qy}, with now the energy $p$ 
playing a similar role as the mass $m$. In the OSEFT there is however an additional term
proportional to $iv\cdot D- \tilde{v}\cdot D$, which is absent in NRQED in the second order correction in the mass
expansion.

At order $1/p^3$ a new local field redefinition eliminates the temporal derivatives
at that order. Thus, after redefining
\noindent 
\begin{equation}
\chi_{v}\rightarrow\chi_{v}^{\prime\prime}=\left(1-\frac{i}{8p^{3}}\slashed{D}_{\perp}\left[\tilde{v}\cdot D\,,\,\slashed{D}_{\perp}\right]+ \frac{i}{16 p^{3}}\slashed{D}_{\perp}^{2}
\left(v\cdot D-\tilde{v}\cdot D\right)
-\frac{i}{16p^{3}}\slashed{D}_{\perp}^{2}\tilde{v}\cdot D\right)\chi_{v}^{\prime}\,\,,
\end{equation}
 one gets
\begin{eqnarray}
\label{Lan-3}
\mathcal{L}^{(3)}_{p, v} & = & \frac{1}{8p^{3}} \chi_{v}^{\prime\prime\dagger}\left\{ \slashed{D}_{\perp}^{4}+\left[\slashed{D}_{\perp},i\tilde{v}\cdot D\right]^{2}-(iv\cdot D-i\tilde{v}\cdot D)\slashed{D}_{\perp}^{2}(iv\cdot D-i\tilde{v}\cdot D)\right\} \chi_{v}^{\prime\prime}\label{eq:L3}\\
 & + & \frac{1}{8p^{3}} \chi_{v}^{\prime\prime\dagger}\left\{ (iv\cdot D-i\tilde{v}\cdot D)\slashed{D}_{\perp}\,\left[i\tilde{v}\cdot D,\slashed{D}_{\perp}\right]-\left[i\tilde{v}\cdot D,\slashed{D}_{\perp}\right]\slashed{D}_{\perp}(iv\cdot D-i\tilde{v}\cdot D)\right\} \chi_{v}^{\prime\prime} \ ,\nonumber 
\end{eqnarray}
with no dependence on temporal derivatives.
Similar local field redefinitions 
could be done at higher orders in the energy expansion.

\subsection{ Propagators of the OSEFT in a thermal bath}
\label{sec:propagators}

In this manuscript 
we will carry out computations of thermal contributions to the polarization tensor in the real time formalism (RTF), as then it is natural to split
the four momentum into an on-shell and off-shell part. In the imaginary time formalism (ITF), where the energies are written in terms of  quantized Matsubara frequencies, such
an splitting cannot be naturally performed. In order to present in full coherence the derivation of the fermion propagators and Feynman diagrams in the theory, we will work in 
the Keldysh formulation of the RTF, see Ref.\cite{Chou:1984es}. However, {\it a posteriori} it is easy to realize how similar 
computations can as well be performed using the ITF.  We defer a discussion on how those  computations should be carried out to 
Appendix~\ref{SecIFT}.

In the Keldysh representation of the RTF  the propagators are formulated as  $2 \times 2$ matrices, in the space
spanned by particle/thermal ghosts. The fermion propagator associated with the lowest order Lagrangian $ \mathcal{L}^{(0)}_{p, v}$ reads  
\begin{equation}
\label{basicProp}
S(k)=P_v \gamma_0  \left [ \left(\begin{array}{cc}
\frac{1}{ v \cdot k +i\epsilon} & 0\\
0 & - \frac{1}{v \cdot k -i\epsilon}
\end{array}\right)+2\pi i\delta(v \cdot k  )\left(\begin{array}{cc}
n_{f}(p + k_0) & n_{f}(p+k_0)\\
-1+n_{f}(p + k_0) & n_{f}(p+k_0)
\end{array}\right)\right]\, ,
\end{equation}
where $n_f(x)= 1/(\exp( |x|/T)+1)$ is the Fermi-Dirac thermal distribution function.

Eq.~(\ref{basicProp}) can be deduced in two different ways. The first way is to start with the Dirac fermion propagator with dependence on the full momentum $q^\mu$, perform
the splitting of Eq.~(\ref{eq:Qpart}),  keeping only the leading terms in a large $p$ expansion.
Alternatively,  one can deduce it from the OSEFT Lagrangian, but realizing that $p$ acts as a sort
of chemical potential for the quantum fluctuations. This last observation 
becomes apparent when we write the Hamiltonian of the full theory in terms of $\chi_{v}$, $\xi_{\tilde v}$, and their canonical momenta. 
At lowest order in the energy expansion it reads
\be
H  = \sum_{p, {\bf v}}  \left(-
p \, \pi_{v} \chi_{v}+ p\, {\tilde \pi}_{\tilde v} \xi_{\tilde v}+ {\cal H}^{(0)}_{p {\bf,v}} \right)  \ ,
\label{eq:HamiltonianMu-1}
\ee
where the fields
\begin{equation}
\pi_{v}  =  \frac{\partial \mathcal{L}^{(0)}_{p, {\bf v}}}{\partial(\partial^{0}\chi_{v})}=i\chi_{ v}^{\dagger}\,,\qquad
{\tilde \pi_{\tilde v}}  =  \frac{\partial\mathcal{L}^{(0)}_{p, {\bf v}}}{\partial(\partial^{0}\xi_{\tilde v})}=i \xi_{\tilde v}^{\dagger}\,.
\end{equation}
 are the canonical conjugate fields of the $\chi_{v}$ and $\xi_{\tilde v}$ fields, respectively, and
\be
{\cal H}^{(0)}_{p {\bf,v}} =\pi_{v} \,\partial_{0}\chi_{v}+\tilde{\pi}_{\tilde v} \,\partial_{0}\xi_{\tilde v}-\mathcal{L}^{(0)}_{p,{\bf v}}\, ,
\ee
 is the Hamiltonian of the OSEFT.

  At every order in the $1/p$ expansion
the propagator is modified, a property that must be taken into account when performing loop computations at a given order in the energy expansion.
In the remaining part of this manuscript we  will use rather the 
retarded, advanced and symmetric particle propagators, which can be constructed in the Keldysh formalism 
in the standard way \cite{Chou:1984es}
\begin{equation}
S_{R/A}  =  S_{11}-S_{12/21}\,,\qquad
S_{S}  =  S_{11}+S_{22}\,.
\end{equation}
The propagators in the Keldysh formalism as derived from considering the  OSEFT Lagrangian up to order $n$ in the energy expansion read
\begin{eqnarray}
\label{RAF-propa}
S^{R/A}(k) & = &  \frac{ P_v \gamma_0   }{k_0  \pm i \epsilon - f({\bf k})} ,\\
S^{S}(k)  & = & P_v \gamma_0 \left( - 2\pi i \delta( k_0 - f({\bf k})) \left( 1 - 2n_f(p +k_0)  \right) \right)  \, .
\end{eqnarray}
The expansion of f({\bf k}) at order $n$  will be denoted as $f^{(n)}({\bf k})$. At lowest order
\be
f^{(0)}({\bf k}) =   k_\parallel \ ,
\ee
and we have defined $ k_\parallel = {\bf k} \cdot {\bf v}$, while 
\be
f^{(1)}({\bf k}) = k_\parallel + \frac{{\bf k}_\perp^2}{2 p} \ , \qquad
 f^{(2)}({\bf k}) = k_\parallel + \frac{{\bf k}_\perp^2}{2 p} - \frac{k_\parallel {\bf k}_\perp^2}{2 p^2 } \ ,
 \label{displaw-2}
  \ee
as follows from Eqs.~(\ref{Lan-1}) and (\ref{Lan-2}), respectively. Note that, for convenience, we keep the propagators above unexpanded even though $f({\bf k})$ contains subleading pieces in  $1/p$.

The propagators for the antiparticle quantum fluctuations can be
also be easily deduced. They  read
\begin{eqnarray}
\label{RAFanti-propa}
{\widetilde S}^{R/A}(k) & = &  \frac{ P_{\tilde v} \gamma_0   }{k_0  \pm i \epsilon - {\tilde f}({\bf k})} ,\\
{\widetilde S}^{S}(k)  & = & - P_{\tilde v} \gamma_0 \left( - 2\pi i \delta( k_0 - {\tilde f}({\bf k})) \left( 1 - 2n_f(-p +k_0)  \right) \right)  \, ,
\end{eqnarray}
where the function ${\tilde f}({\bf k})$ can be obtained from $f({\bf k})$, with the replacements ${\bf v} \rightarrow - {\bf v}$ and $p  \rightarrow -p$.
Note the extra minus sign in the symmetric antiparticle propagator, absent in its particle counterpart. The presence of this additional minus sign might be deduced
from the full theory.

When performing computations of Feynman diagrams at a given order in the $1/p$ expansion, the above propagators should eventually be Taylor expanded,
assuming that $k_0, k \ll p$. However, in practice, it is more convenient to carry out the $k_0$ integral before performing these expansions.
We will denote the pieces of this expansion as $S_{(n)}$, where $n$ labels the order of the expansion. Note
also the distribution function is the  symmetric propagator must also be expanded in $k_0$.


Also note that due to the local field redefinitions  introduced beyond leading order  the propagators  deduced from the OSEFT and those derived from the full theory
also differ beyond leading order. However, the dispersion relations coincide, as they should.

\section{Computation of the retarded photon polarization tensor for soft momentum }
\label{sec:polarization}

In this section we compute in the framework of the OSEFT  the one-loop  retarded photon polarization tensor up  to third order in the energy expansion, assuming
that the photon momentum $l$ is soft, or of order $eT$. In a thermal plasma it is well-known that the leading order behavior
is given by the HTL polarization tensor \cite{Braaten:1989mz,Frenkel:1989br} (see also Ref.~\cite{Carrington:1997sq} for an alternative derivation using the
RTF).  As it is known that the HTLs are dominated by the contribution
of almost on-shell particles and antiparticles with energies $\sim T$, we will then assume $ p \sim T$.  We will also assume that
 $ l_\mu \ll p$, but $ l_\mu \sim k_\mu $.  
We will effectively show that the OSEFT allows to reproduce, to leading order, the HTL polarization tensor, but also allows us to extract in a very systematic way subleading corrections to the HTLs.

\begin{figure}
\label{fig}
\centering
\includegraphics{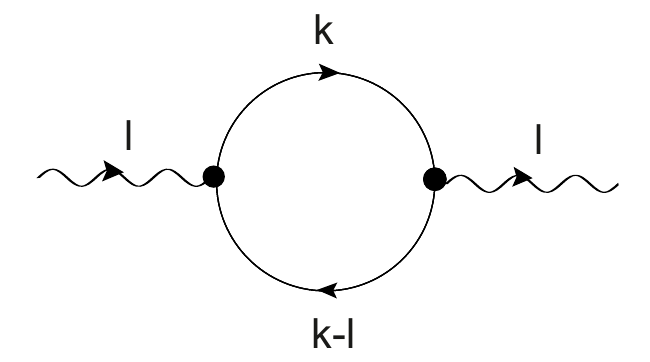}
\vspace{-0.5cm}\includegraphics{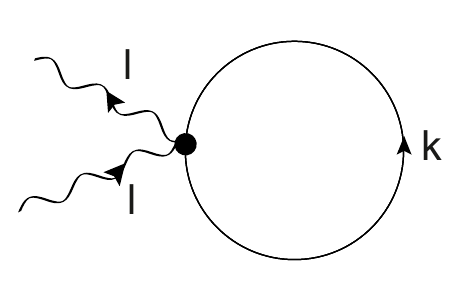}
\caption
[]{We display the two topologies that contribute to the photon self-energy at one loop in the OSEFT. The blob symbolizes any vertex that may contribute to a given order.}
\end{figure}

There are two topologically different kind of diagrams that contribute at one-loop to the photon polarization tensor in the OSEFT, that we call bubble and tadpole diagrams,
respectively.  
The generic form of the particle's contribution  to the retarded polarization tensor for the bubble diagrams reads (see Fig.~\ref{fig} left)
\begin{eqnarray}
\Pi^{\mu\nu}_{\rm b}(l) &=&- \frac{i}{2} \sum_{p, {\bf v}}\int\frac{d^{4} k }{(2\pi)^{4}} \Big ({\rm Tr}[V^{\mu}\, S_{S}(k-l)\, V^{\nu}\, S_{R}(k)]+{\rm Tr}[V^{\mu}\, S_{A}(k-l)\, V^{\nu}\, S_{S}(k)]\Big )\nonumber  \\
& - &   \frac{i}{2} \sum_{p, {\bf v}}\int\frac{d^{4} k }{(2\pi)^{4}} 
\Big ({\rm Tr}[V^{\mu}S_{A}(k-l)V^{\nu}S_{A}(k)]+{\rm Tr}[V^{\mu}S_{R}(k-l)V^{\nu}S_{R}(k)] \Big) \ ,
\label{eq:bubble}
\end{eqnarray} 
while the tadpole diagrams are expressed as (see Fig.~\ref{fig} right)
\begin{equation}
\Pi^{\mu\nu}_{\rm t}(l) = \frac{i}{2} \sum_{p, {\bf v}}\int\frac{d^{4} k }{(2\pi)^{4}} \Big ({\rm Tr}[W^{\mu \nu }\, (S_{S}(k) + S_{R}(k) + S_{A}(k))]\Big )
\ ,
\label{eq:tad}
\end{equation}
where we have dropped all the subindices that label the order of the energy expansion in the  vertices $V^\mu$, $W^{\mu \nu}$, and in the propagators. The $k$ and $l$ dependence of $V^\mu$ and $W^{\mu \nu}$ must be understood. We have a single sum on $p$ and {\bf v} above because the interaction with soft photons cannot change $p$ and {\bf v}. Further, we have taken into account that 
no vertex connects particle and thermal ghost propagators (${\it i.e.}$, $V^\mu_{12/21} = W^{\mu \nu}_{12/21} =0$).

The appearance of the tadpole diagrams in the effective field theory, which are absent in the full theory, 
take into account particle-photon interactions mediated by an off-shell antiparticle (or viceversa for the antiparticle-photon interactions).
We will see that they are necessary in order to 
fulfil the Ward identity $l_\mu \Pi^{\mu \nu} =0$  of the fundamental theory at every order in the energy expansion.

Some generic simplifications occur in the computation of both the bubble and tadpole Feynman diagrams. First, one notices that 
 the last line of  Eq.~(\ref{eq:bubble}) vanishes, as one can immediately check after performing the
$k_0$ integral. This is due to the fact that the poles of the two retarded (or advanced) propagators lie on the same side of the complex plane. Similarly, one can check that terms
proportional to the retarded and the advanced fermion propagator vanish in Eq.~(\ref{eq:tad}).

The non-vanishing terms of the bubble contribution to the retarded polarization tensor can be computed in a rather systematic and compact way thanks to the local field redefinitions introduced in Sec.~\ref{sec:Lag}. Feynman rules associated with the photon-fermion interactions can be extracted at every order in the $1/p$ expansion from the Lagrangians written down in Subsec.~\ref{sec:Lag}.
The corresponding vertex appearing in the bubble diagram at order $n$ is denoted by  $V^\mu_{(n)}$, and for completeness we present in Table \ref{FeynmanR_V} explicit values of those vertices for $n=0,1,2$. The evaluation of the bubble diagrams then requires the computation of different traces, that in order to simplify the notation we denote  as
 \be
 {\widehat {\rm  Tr}}[V_{(n)}^{\mu}\, V_{(m)}^{\nu}]  \equiv {\rm Tr}[P_v \gamma_0 V_{(n)}^{\mu}\,P_v  \gamma_0  V_{(m)}^{\nu}] 
\ee
for the particle fluctuations. In the bubble diagrams, it turns out to be convenient to defer the expansion
in 1/p of the propagators. 
Hence we use the general form of the symmetric,
retarded and advanced propagators, perform the $k_0$ integral, and then expand the result in $1/p$  
at the desired order. 
The $k_0$ integral that appears in all the bubble diagrams can be easily performed,
given the form 
of the propagators in the Keldysh representation, see Eq.~(\ref{RAF-propa}), and also due to the fact that there is no frequency dependence in the vertices of the theory.
The energy integral present in all the bubble  diagrams is of the form
\begin{eqnarray}
I_{k_{0}} & = & \left(-2\pi i\right)\frac{1}{2}\int\frac{dk_{0}}{2\pi}\left[\left(1-2n_{f}(p+k_{0}-l_{0})\right)\delta\left(k_{0}-l_{0}-f({\bf k-l})\right)\frac{1}{k_{0}-f({\bf k})+i\epsilon}\right]\nonumber \\
 & + & \left(-2\pi i\right)\frac{1}{2}\int\frac{dk_{0}}{2\pi}\left[\left(1-2n_{f}(p+k_{0})\right)\delta\left(k_{0}-f({\bf k})\right)\frac{1}{k_{0}-l_{0}-f({\bf k-l})-i\epsilon}\right]\nonumber \\
\nonumber \\
 & = & \frac{-i}{l_{0}+f({\bf k-l})-f({\bf k})+i\epsilon}\left[n_{f}(p+f({\bf k}))-n_{f}(p+f({\bf k-l}))\right] \ .
 \label{eq:IntExp}
\end{eqnarray}
Reaching this compact formula was the main reason why we kept the propagators in (\ref{RAF-propa}) and (\ref{RAFanti-propa}) unexpanded.

Thus, if one wants to compute the bubble diagram at a given order in $1/p$ one simply has to expand for large $p$ an integral like the one above, in addition to considering the possible $p$ dependence of the vertices of the diagram. In appendix~\ref{System-app} we present the result of these expansions as soon as the fermion dispersion law is fixed at order $1/p^2$, that is,
we present the explicit values of $I_{k_{0}}^{(1)}, I_{k_{0}}^{(2)}$ and $I_{k_{0}}^{(3)}$. Note that $I_{k_{0}}^{(0)} =0$,    
due to the fact that Eq.~(\ref{eq:IntExp})  depends on a difference of the fermion distribution functions, and  $l_\mu, k_\mu \ll p$. If
 we consider the contribution to a bubble
diagram with  propagators  at order $n$, the bubble diagram is at least  one order higher in the counting, that is, it is at least of order $n+1$ in the $1/p$ expansion.

The tadpole  diagrams are very easily computed, and basically only require the knowledge of the vertices $W^{\mu \nu}_{(n)}$. 
However, starting at order $1/p^3$ they become ambiguous, even after regulating the ultraviolet (UV) divergence that appears at that order. The ambiguity amounts to local counterterms built out of the electromagnetic stress tensor, see Eq.~(\ref{c.t.}),
 and hence it is innocuous for the consistency of the OSEFT. Nevertheless, a clear prescription must be given in order to display reproducible results.
We present in Table \ref{FeynmanR_W} 
the value of the traces of the particle projectors times the vertices required in the computation of the polarization tensor  presented in this manuscript. Since the ordering of the fields will be relevant to discuss the ambiguity,  in Table \ref{FeynmanR_W} we present the results in the case that the photon with incoming momentum is to the left of the photon with outgoing momentum only. The opposite case is obtained by just changing the sign of $l$.

Note that to the above particle's tadpole and bubble contributions one
 should also include analogous antiparticles' contributions, ${\widetilde \Pi^{\mu\nu}}_{\rm b}(l)$ and ${\widetilde \Pi}^{\mu\nu}_{\rm t}(l)$, which are similarly computed 
 using the corresponding antifermion propagators and vertices.  In particular, to simplify the notation we denote
\be
 {\widehat {\rm  Tr}}[{\widetilde V}_{(n)}^{\mu}\, {\widetilde V}_{(m)}^{\nu}]  \equiv {\rm Tr}[P_{\tilde v} \gamma_0 {\widetilde V}_{(n)}^{\mu}\,P_{\tilde v}  \gamma_0 {\widetilde V}_{(m)}^{\nu}] 
\ee
the required traces for the antiparticle fluctuations. For the antiparticle fluctuations we denote the same integral that appears in Eq.~(\ref{eq:IntExp}) by ${\tilde I}_{k_0}$.

\begin{table}
\begin{tabular}{c c l}
\hline
$V_{(0)}^{\mu}$ & $ =$ & $e\gamma_{0}v^{\mu} $
\\
$V_{(1)}^{\mu}$ & $ =$ & $ \frac{e}{p} \gamma_{0}\left[\left(k_{\perp}^{\mu}+\frac{1}{2}l_{\perp}^{\mu}\right)-\frac{i}{2}\sigma_{\perp}^{\mu\alpha}l_{\alpha}\right]$%
 \\
$V_{(2)}^{\mu} $ & $= $&$ -\frac{e}{4p^{2}}\gamma_{0}\left[\left(l_{\parallel}+2k_{\parallel}\right)l_{\perp}^{\mu}+2\left(l_{\parallel}+2k_{\parallel}\right)k_{\perp}^{\mu}+ \left({\bf l}_{\perp}^{2}+ 2{\bf l}_{\perp}\cdot{\bf k}_{\perp}+2{\bf k}_{\perp}^{2}\right)\delta^{\mu i}v^{i} 
+ \frac{1}{2}\left(\left(\tilde{v}\cdot l\right)l_{\perp}^{\mu} +{\bf l}_{\perp}^{2}\tilde{v}^{\mu}\right)\right] $ \\
\hline
\end{tabular}
\caption{Feynman rules for vertices involving one photon line at different orders in the energy expansion. These are derived from the Lagrangians of Eqs.~(\ref{Lan-0}), (\ref{Lan-1}) and (\ref{Lan-2}), respectively.
The momentum carried out by the incoming photon is $l^\mu$, while $k^\mu$ is the momentum of the incoming fermion.  
We have ignored in
$V_{(2)}^{\mu}$ a spin-dependent contribution, as it does not contribute to the bubble diagram at the order considered here. The associated Feynman rules for the antiparticles,
${\widetilde V}^\mu_{(n)}$,
are deduced from those of the particles, performing the change $p \rightarrow -p$ and ${\bf v} \rightarrow -{\bf v}$. }
\label{FeynmanR_V}
\end{table}
In the sequel we will present the result of the retarded polarization tensor up to $1/p^3$ order, stressing again that to zero order it vanishes, $\Pi^{\mu \nu}_{(0)} = 0$.

\begin{table}
\begin{tabular}{c c l}
\hline
${\rm Tr}( P_v  \gamma_0W_{<(1)}^{\mu\nu})$ & $=$  & $   \frac{ e^{2}}{p} P_{\perp}^{\mu\nu} $ 
\\
$ {\rm Tr}( P_v  \gamma_0 W_{<(2)}^{\mu\nu}(k,l) )$ & $=$ & $ \frac{e^{2}}{2 p^{2}}\left[-2k_{\parallel}P_{\perp}^{\mu\nu} + \left( (-l^\mu_\perp + 2 k^\mu_\perp) \delta^{\nu i} + (-l^\nu_\perp + 2 k^\nu_\perp) \delta^{\mu i} \right) v^i\right]
$
 \\
$ {\rm Tr}( P_v  \gamma_0 W_{<(3)}^{\mu\nu}(k,l)) $ & $= $ & $\frac{1}{4p^{3}}\left[ P_{\perp}^{\mu\nu}\left(-2\boldsymbol{k}_{\perp}^{2}-\boldsymbol{l}_{\perp}^{2}+4k_{\parallel}^{2}
+(\tilde{v}\cdot l)^{2}-4\tilde{v}\cdot l k_{\parallel}\right)+4k_{\perp}^{\mu}k_{\perp}^{\nu}
-\tilde{v}^{\mu}\tilde{v}^{\nu}\boldsymbol{l}_{\perp}^{2}\right. $\\
& & $\left. -4v^{i}v^{j}\delta^{\mu i}\delta^{\nu j}(\boldsymbol{l}_{\perp}^{2}+\boldsymbol{k}_{\perp}^{2}-2\boldsymbol{l}_{\perp}\boldsymbol{k}_{\perp})\right] $ \\
$ $ & $+ $ &$ \frac{1}{4p^{3}}\left[-\tilde{v}^{\mu}l_{\perp}^{\nu}(\tilde{v}\cdot l-2k_{\parallel})-(8k_{\parallel}-2\tilde{v}\cdot l)k_{\perp}^{\mu}\delta^{\nu i}v^{i}-2\delta^{\mu i}v^{i}\left((\boldsymbol{l}_{\perp}^{2}-\boldsymbol{l}_{\perp}\boldsymbol{k}_{\perp})\tilde{v}^{\nu} \right.\right. $\\
& & $\left.\left. +(\tilde{v}\cdot l-k_{\parallel})l_{\perp}^{\nu}\right)-2k_{\perp}^{\mu}l_{\perp}^{\nu}+(\mu\leftrightarrow\nu)\right]  $\,\\
\hline
\hline
$\quad\quad\quad W_{>(n)}^{\mu\nu}(k,l)$ & $=$ & $W_{<(n)}^{\mu\nu}(k,-l)$\\
$\quad\quad\quad W_{(n)}^{\mu\nu}(k,l)$ & $=$ & $W_{<(n)}^{\mu\nu}(k,l)+W_{>(n)}^{\mu\nu}(k,l)$\\
\hline
\end{tabular}
\caption{ Traces needed for the computation of the tadpole-like diagrams for the particle sector. The vertices $W^{\mu \nu}_{(n)}$  involve two photon lines and are computed from Eqs.~(\ref{Lan-1}), (\ref{Lan-2}) and (\ref{Lan-3}), for the cases $n=1,2$, and $3$ respectively. The momentum carried out by the incomming photon is $l^\mu$, while $k^\mu$ is the momentum of the incoming fermion. The $<$ subscript means that only the contributions corresponding to the incoming momentum carried out by the left photon are displayed. The contributions corresponding to the incoming momentum carried out by the right photon, which will be labeled by the $>$ subscript, may be obtained by changing the sign of $l$, as displayed in the second-to-last line. 
The full expression reads $W_{(n)}^{\mu\nu}(k,l)$ and  is displayed in the last line.
Note that for $n=1$ there is no dependence on $l$ or $k$ and hence we drop them from the expressions. The corresponding expressions for the antiparticle sector may be obtained by changing $v\to {\tilde v}$ and $p\to -p$. 
}
\label{FeynmanR_W}
\end{table}

\subsection{ Polarization tensor at order $e^2 T^2$}
\label{Orderp1}

 We start by computing the retarded polarization tensor at the  first non-trivial order in the energy expansion.
The bubble diagram can be immediately evaluated, and after performing the $k_0$ integral as prescribed in Eq.~(\ref{eq:IntExp}), it  reads
\begin{equation}
\Pi_{b, (1)}^{\mu\nu}(l) +{\widetilde \Pi}_{b, (1)}^{\mu\nu}(l)=-i \sum_{p, {\bf v}}\int\frac{d^{3}{\bf k}}{(2\pi)^{3}}\left( \widehat {\rm  Tr}[V_{(0)}^{\mu}\, V_{(0)}^{\nu}]
\, I_{k_{0}}^{(1)} - \widehat {\rm  Tr}[{\widetilde V}_{(0)}^{\mu}\, {\widetilde V}_{(0)}^{\nu}] \, {\tilde I}_{k_{0}}^{(1)}  \right)
\,,
\end{equation}
where the explicit value of $I_{k_{0}}^{(1)}$  and $ {\tilde I}_{k_{0}}^{(1)}$ can be found in Appendix A, see Eq.~(\ref{Ik-1}). 
Note that the antiparticles contribute with a relative minus sign compared to the particle's contribution, due to the form of the antiparticle symmetric propagator.
We then reach to 
\begin{equation}
\Pi_{b, (1)}^{\mu\nu}(l) +{\widetilde \Pi}_{b, (1)}^{\mu\nu}(l)=-2e^{2} \sum_{p, {\bf v}}\int\frac{d^{3}{\bf k}}{(2\pi)^{3}}  \frac{d n_f}{dp} l_{\parallel}
\left(\frac{v^{\mu}v^{\nu}}{v\cdot l} - \frac{{\tilde v}^{\mu}{\tilde v}^{\nu}}{{\tilde v}\cdot l} \right) \label{HTLb}
 \ ,
\end{equation}
where for the retarded  boundary conditions $l_0 \rightarrow l_0 + i \epsilon$. 
The tadpole diagram contribution at this order is expressed as
\begin{eqnarray}
\Pi^{\mu\nu}_{\rm t ,(1)}(l) +{\widetilde \Pi}_{t, (1)}^{\mu\nu}(l)&= & \frac{i}{2} \sum_{p, {\bf v}}\int\frac{d^{4} k }{(2\pi)^{4}} {\rm Tr}[W_{(1)}^{\mu \nu }\, S_{S}^{(0)}(k) +
{\widetilde W}_{(1)}^{\mu \nu }\, {\widetilde S}_{S}^{(0)}(k) ] \ ,
\label{tp-1}
\end{eqnarray}
where the required traces at this order needed for the computation can be read in Table~\ref{FeynmanR_W}. More explicitly, one finds

\begin{eqnarray}
\nonumber
\Pi_{t, (1)}^{\mu\nu}(l)+{\widetilde \Pi}_{t, (1)}^{\mu\nu}(l) &= &- ie^{2}\sum_{p, {\bf v}}\int\frac{d^{4}k}{(2\pi)^{4}} \frac{ P^{\mu \nu}_\perp}{p} (2\pi i)  \Big (  \delta( v \cdot k) (1-2 n_f(p ))  
\\
& + & \delta( {\tilde v} \cdot k) (1-2 n_f( -p )) \Big)  \ .
\end{eqnarray}
Note that the relative minus sign between the particle and antiparticle symmetric propagators is compensated here by the relative minus sign in the corresponding
vertex for the tadpole diagram.
After performing the integral on $k_0$ we end up with
\begin{equation}
\label{tadpole-1}
\Pi_{t, (1)}^{\mu\nu}(l)+{\widetilde \Pi}_{t, (1)}^{\mu\nu}(l)= 2 e^{2}\sum_{p, {\bf v}}\int \frac{d^{3}{\bf k}}{(2\pi)^{3}}
\frac{ P^{\mu \nu}_\perp}{p} (1-2  n_f(p ) )  \ .
\end{equation}

We need now to give a precise meaning to $\sum_{p, {\bf v}}$ in Eq.~(\ref{HTLb}) and Eq.~(\ref{tadpole-1}). Recall that $\sum_{p, {\bf v}}$ together with $\int\frac{d^{3}{\bf k}}{(2\pi)^{3}}$ arise from the splitting of a single
variable ${\bf q}$ in a large component $p{\bf v}$ and a residual momentum ${\bf k}$.
We should be able to re-express the above integrands in terms of the full momemtum
${\bf q}$, see Eq.~({\ref{eq:Qpart}), as this is the variable used in the full theory computations.  If we define the quantities
$k_{\parallel,{\bf q}} \equiv {\bf k} \cdot { \bf{\hat q}}$, where ${  \bf{\hat q}}=\frac{\bf q}{q}$, $q=\vert {\bf q}\vert$, 
and ${\bf k}_{\perp,{\bf q}} \equiv{\bf k} - {  \bf{\hat q}} k_{\parallel,{\bf q}}$,
then one has to take into account that
\begin{eqnarray}
\label{p-back}
p &=& q - k_{\parallel,{\bf q}} + \frac{{\bf k}_{\perp,{\bf q}}^2 }{2 q} +{\cal O}(\frac{1}{q^2}) \ ,\\
\label{v-back}
{\bf v} & = & {  \bf{\hat q}} -  \frac{{\bf k}_{\perp,{\bf q}} }{q}  - \frac{  {\bf \hat q} {\bf k}_{\perp,{\bf q}}^2 + 2  k_{\parallel,{\bf q}} {\bf k}_{\perp,{\bf q}}}{2 q^2}+{\cal O}(\frac{1}{q^3})
\ , \\
\label{n-back}
n_f(p) &= & n_f(q) +  \frac{d n_f}{dq} \left( -  k_\parallel^{\bf q}  + \frac{{\bf k}_{\perp,{\bf q}}^2 }{2 q} \right) + \frac 12 \frac{d^2 n_f}{dq^2}   k_{\parallel, {\bf q}}^2 +
{\cal O}(\frac{1}{q^3}) \ .
\end{eqnarray}
We then use  the identification (see Ref.~\cite{Luke:1999kz})
\begin{equation}
 \sum_{p, {\bf v}}\int\frac{d^{3}{\bf k}}{(2\pi)^{3}} \equiv \int\frac{d^{3}{\bf q}}{(2\pi)^{3}} \ .
\label{def}
\end{equation} 
 At this point one notes that the $T=0$ contribution to the tadpole is UV divergent. We regulate such a divergence in dimensional regularization (DR),
with $d=3 + \epsilon$, which puts the $T=0$ contribution to zero. Then, after adding the bubble and tadpole contributions, which are needed in order
that the resulting tensor respects the Ward identity, 
 we reach to the 
result 
\begin{equation}
\label{fullPi-1}
\Pi_{ {\rm total}, (1)}^{\mu\nu}(l) =4  e^2 \int\frac{d^{3}{\bf q}}{(2\pi)^{3}} \left \{\frac{d n_f}{dq} \left( \delta^{\mu 0}\delta^{\nu 0} - l_0\frac{v^\mu_{\bf q} v^\nu_{\bf q} }{ v_{\bf q}\cdot l  } \right) + {\cal O}(\frac{1}{q^2}) \right\} \ ,
\end{equation}
where we have performed an integration by parts of $n_f$, and
we have defined $v^\mu_{\bf q} \equiv ( 1, {\bf \hat q})$. We have also performed a change of variables in the contribution coming from the antiparticles,
 ${\bf v}_{\bf q}  \rightarrow - {\bf v}_{\bf q}$, such that the antiparticle contribution can be written in
the same form as the particle contribution. 
In this way we reproduce
to leading order the  HTL polarization tensor. Note that in the result shown above we have neglected corrections of
order $1/q^2$. Those pieces turn out to be important when we compute higher order corrections to the polarization tensor, and will be discussed  in Appendix ~\ref{Cancellk2}.

For completeness, we present the explicit form of the HTL polarization tensor, that can be found out after performing the angular integrals of Eq.~(\ref{fullPi-1}). More explicitly 
%
\begin{eqnarray}
 \Pi_{{\rm total},(1)}^{00} (l_0, {\bf l}) & = &\Pi^L_{{\rm total},(1)}
(l_0, {\bf l}) \ ,\\[2mm]
 \Pi^{0i}_{{\rm total},(1)} (l_0, {\bf l}) & = & 
l_0 \, \frac{l^i}{|{\bf l}|^2}\, \Pi^{L}_{{\rm total},(1)}(l_0, {\bf l}) \ , \\
\Pi^{ij}_{ {\rm total},(1)} (l_0, {\bf l}) & = &  \left[
 \left ( \delta^{ij}- \frac{l^i l^j}{|{\bf l}|^2} \right)
 \Pi^T_{{\rm total},(1)} (l_0,{\bf l})+ \frac{l^i l^j} {|{\bf l}|^2} \, \frac{l_0^2 }{|{\bf l}|^2} \,
\Pi^L_{{\rm total},(1)} (l_0, {\bf l}) \right] \ ,
\label{resultreal}
\end{eqnarray}
$\!\!$expressed in terms of the longitudinal and transverse components, given by
\begin{eqnarray}
\label{pipiL}
\Pi^L_{{\rm total},(1)} (l_0, {\bf l}) & = & m^2 _{D} \left( \frac{l_0}{2|{\bf
l}|} \left(
 \,{\rm ln\,}\left|{\frac{l_0+|{\bf l}|}{l_0-|{\bf l}|}}\right| 
-i \pi \, \Theta(|{\bf l}|^2 -l_0^2) \right)
-1  \right) \ , \\
 \Pi^T_{{\rm total},(1)} (l_0, {\bf l}) & = &- m^2 _{D} \, \frac{l_0^2}{2 |{\bf
l}|^2} \left[ 1 + \frac12 \left( \frac{|{\bf l}|}{l_0} -
\frac{l_0}{|{\bf l}|} \right) \, \left( {\rm ln\,} \left|{\frac{l_0+
|{\bf l}|}{l_0-|{\bf l}|}}\right| -i \pi \, \Theta(|{\bf l}|^2 -l_0^2)
 \right) \, \right] \ .
 \label{pipiT}
\end{eqnarray}
respectively. Here $\Theta$ is the step function, and $m^2_D = \frac{e^2 T^2}{6}$ is the Debye mass squared. The imaginary part of the polarization tensor gives account of Landau damping.

\subsection{ Polarization tensor at order $e^2 T$}
 \label{Orderp2}

At this order in the expansion we find a vanishing contribution to the polarization tensor in
the rotational invariant thermal plasma. Let us explain how this happens first for the particle's contribution,
the antiparticle contribution is similarly computed.

Let us consider first the tadpole  diagrams, which read
\begin{eqnarray}
\Pi^{\mu\nu}_{\rm t ,(2)}(l) &= & \frac{i}{2} \sum_{p, {\bf v}}\int\frac{d^{4} k }{(2\pi)^{4}} {\rm Tr}[W_{(2)}^{\mu \nu }(k,l)\, S_{S}^{(0)}(k) + W_{(1)}^{\mu \nu }\, S_{S}^{(1)}(k) ]  .
\label{tp-2}
\end{eqnarray}
The explicit expressions of the  tadpole contributions at this order can be written down after using the values of the traces displayed in Table~\ref{FeynmanR_W}. After
expressing these contributions in terms of the original variables it is not difficult to realize that they cancel 
 after performing the angular integration over  ${\bf \hat q}$ (note that to leading order ${\bf v} \sim {\bf \hat q}$).

A careful inspection 
of all the bubble  diagrams that appear at this order leads to
\begin{equation}
\Pi_{b,(2)}^{\mu\nu}(l) =  -i  \sum_{p, {\bf v}}\int\frac{d^{3}{\bf k}}{(2\pi)^{3}}
\left\{ \left( {\widehat {\rm  Tr}}[V_{(1)}^{\mu}\, V_{(0)}^{\nu}]+  {\widehat {\rm  Tr}}[V_{(0)}^{\mu}\, V_{(1)}^{\nu}] \right)\, I_{k_{0}}^{(1)}+  {\widehat {\rm  Tr}}[V_{(0)}^{\mu}\, V_{(0)}^{\nu}]\, I_{k_{0}}^{(2)}\right\} \,.
\end{equation}
 After using Eqs.~(\ref{Ik-1}) and (\ref{Ik-2}), together with the Feynman rules of Table  \ref{FeynmanR_V}, the above contribution can be expressed as
\begin{eqnarray}
\label{Pibubble-2pre}
\Pi_{b,(2)}^{\mu\nu}(l) & = &  e^{2} \sum_{p, {\bf v}}\int\frac{d^{3}{\bf k}}{(2\pi)^{3}} \Bigg [
   \frac{1}{p}  \frac{d n_f}{dp}  \left\{ l_{\parallel}\left(l_{\perp}^{\mu}v^{\nu}+l_{\perp}^{\nu}v^{\mu}\right)\frac{1}{v\cdot l}+v^{\mu}v^{\nu}\left[\frac{{\bf l}_{\perp}^{2}}{v\cdot l}
+\frac{{\bf l}_{\perp}^{2}l_{\parallel}}{(v\cdot l)^{2}}\right]\right\}  \nonumber \\
&+& \frac{d^2 n_f}{dp^2} v^{\mu}v^{\nu} \frac{l_{\parallel}^2}{(v\cdot l) } \Bigg] \ ,
\label{vp}
\end{eqnarray}
where we have not written terms linear in $k_{\parallel}$ and ${\bf k}_{\perp}$, as they cancel out if we assume that the formal measure of the ${{\bf k}}$-integration is invariant under ${\bf k}\to -{\bf k}$.
We re-express the value of the above integrand in terms of the original variable ${\bf q}$ to reach to
\begin{eqnarray}
\label{Pibubble-2}
\Pi_{b,(2)}^{\mu\nu}(l) &=&  e^{2}  \int\frac{d^{3}{\bf q}}{(2\pi)^{3}} \frac 1q \frac{d n_f}{dq}
  \left \{ l_{\parallel,\bf q}\left(l_{\perp,{\bf q}}^{\mu}v_{\bf q}^{\nu}+{l}_{\perp,{\bf q}}^{\nu}v_{\bf q}^{\mu}\right)\frac{1}{v\cdot l} + v_{\bf q}^{\mu}v_{\bf q}^{\nu}\,
 \left ( \frac{{\bf l}_{\perp,{\bf q}}^{2}-2 l_{\parallel,{\bf q}}^{2}}{v_{\bf q}\cdot l} \right. \right.
   \nonumber \\
 & + & \left. \left. \frac{ {\bf l}_{\perp, {\bf q}}^{2}l_{\parallel,{\bf q}} }{(v_{\bf q}\cdot l)^{2}} \right)  + {\cal O}(\frac{1}{q})  \right\} \ ,
\end{eqnarray}
where we have integrated by parts the fermionic distribution function.
 What it is most surprising, not obvious at first sight, is that Eq.~(\ref{Pibubble-2}) vanishes, 
 after performing the angular integration. 
 
 As the antiparticle contribution at this order also vanishes, we then conclude that there is no finite contribution to the polarization tensor at order $e^2 T$.

\subsection{
Polarization tensor at order $e^2 T^0$}
\label{Orderp3}

We distribute this section in two subsections. In the first one we display the (unambiguous) contribution of the bubble diagram, and in the second one we illustrate the inherent ambiguity of the tadpole contributions at this order by calculating them in two apparently equivalent ways. We shall focus on the contribution of particles, since the contribution of antiparticles may be easily  obtained from it, as it has been done in previous sections.  

\subsubsection{Bubble diagrams}

At order $1/p^3$  the bubble  diagrams contributing to the polarization tensor can be expressed as
\begin{eqnarray}
\Pi_{b,(3)}^{\mu\nu}(l) & = & -i  \sum_{p, {\bf v}}\int\frac{d^{3}{\bf k}}{(2\pi)^{3}}
\left\{ {\widehat {\rm  Tr}}\left[V_{(0)}^{\mu}V_{(0)}^{\nu}\right]I_{k_{0}}^{(3)}+ {\widehat {\rm  Tr}}\left[V_{(0)}^{\mu}V_{(1)}^{\nu}+V_{(1)}^{\mu}V_{(0)}^{\nu}\right]I_{k_{0}}^{(2)}
\right. \nonumber \\
 & + & \left. {\widehat {\rm  Tr}}\left[V_{(1)}^{\mu}V_{(1)}^{\nu}\right]I_{k_{0}}^{(1)}+{\widehat {\rm  Tr}}\left[V_{(2)}^{\mu}V_{(0)}^{\nu}+V_{(0)}^{\mu}V_{(2)}^{\nu}\right]I_{k_{0}}^{(1)}\right\} 
 \ ,
 \label{eq:SEap}
\end{eqnarray}
where the needed values of the $I_{k_0}^{(n)}$ functions can be found in the Appendix \ref{System-app}. We note that these functions depend both linearly and quadratically
on ${\bf k}$. However, such a dependence can be obviated, since the linear terms can be dropped, as argued before,  
while the quadratic terms of  $I_{k_0}^{(3)}$   are canceled if we re-express the
contribution computed at lower orders in the $1/p$ expansion in terms of the full momentum ${\bf q}$. A proof of how this happens for the tadpole contribution is presented in Appendix \ref{Cancellk2}.

With the basic rules already explained on how to express the OSEFT loop integrals in terms of the full momentum ${\bf q}$ we then reach to
\begin{eqnarray}
\label{Pipi-3b}
\Pi_{b,(3)}^{\mu\nu}(l) &=&  - 2 e^{2}  \int\frac{d^{3}{\bf q}}{(2\pi)^{3}} \frac {1}{q^2}\frac{d n_f}{dq}  \left \{
\left[\frac 13 \frac{ l_{\parallel}^{3}- 3{{\bf l}}_{\perp}^{2}l_{\parallel} }{v\cdot l}+\frac{1}{4}\frac{{{\bf l}}_{\perp}^{4}-3{{\bf l}}_{\perp}^{2}l_{\parallel}^{2}}{\left(v\cdot l\right)^{2}}+\frac{1}{4}\frac{{\bf l}_{\perp}^{4}l_{\parallel}}{\left(v\cdot l\right)^{3}}\right] \, v^{\mu}v^{\nu} \right.   \nonumber \\
 & + &   \left. \frac{1}{4}\left[\frac{{\bf l}_{\perp}^{2}- 2l_{\parallel}^{2} }{v\cdot l}+\frac{{\bf l}_{\perp}^{2}l_{\parallel}}{\left(v\cdot l\right)^{2}} - \frac 12 \frac{l_{\parallel}(\tilde{v}\cdot l) }{v\cdot l}\right]\,\left(v^{\mu}l_{\perp}^{\nu}+v^{\nu}l_{\perp}^{\mu}\right)  -\frac{1}{4} \frac{ l_{\parallel}{\bf l}_{\perp}^{2}}{v\cdot l}\, P_{\perp}^{\mu\nu}  \right. 
 \nonumber \\
 &-& \left. \frac 18 \frac{{\bf l}_{\perp}^{2}l_{\parallel}}{v\cdot l} \left(\tilde{v}^{\mu}v^{\nu}+\tilde{v}^{\nu}v^{\mu}\right) - \frac 14 \frac{{\bf l}_{\perp}^{2}l_{\parallel}
 }{ v \cdot l}
 \left(\delta^{i \mu}v^{\nu}+\delta^{i \nu}v^{\mu}\right) v^i  + {\cal O}(\frac{1}{q})   \right\} \ ,
\label{b3}
\end{eqnarray}
where, as in previous orders, 
 we have carried out an integration by parts in some terms in order to have the first derivative of the distribution function in all of them. Note that in order not to overcharge the notation  we have dropped the subindex ${\bf q}$ in all the variables of the integrand above, that should be understood.

We  note that the bubble contribution alone, as it happens at order $1/p$, does not fulfil the Ward identity of QED. From Eq.~(\ref{b3}), it is easy to see that $l_\mu \Pi_{b,(3)}^{\mu\nu}(l)$ contains only local terms that lead to
 \be
l_\mu \Pi_{b,(3)}^{\mu i}(l_{0},{\bf l}) =-\frac{e^2{\bf l}^2l^i}{60\pi^2} \neq 0 \, .
\label{wid3}
\ee
 This can also be checked by explicitly performing the angular integrals in Eq.~(\ref{b3}). 
 One finds
{\begin{equation}
\Pi_{b,(3)}^{00}(l_{0},{\bf l})= \frac{e^{2}}{ 144 \pi^2 }\left[16{\bf l}^{2}-6l_{0}^{2}+3\frac{l_{0}}{|{\bf l}|}\left(l_{0}^{2}-3{\bf l}^{2}\right)  {\rm ln\,} \left ({\frac{l_0+
|{\bf l}|}{l_0-|{\bf l}|}}\right)
 \right] \ ,
\end{equation}
and 
\begin{equation}
\Pi_{b,(3)}^{0i}(l_{0},{\bf l}) = \frac{l_0 l^i}{|{\bf l}|^2}  \Pi_{b,(3)}^{00}(l_{0},{\bf l}) \ ,
\end{equation}
so that $l_\mu \Pi_{b,(3)}^{\mu 0}(l_{0},{\bf l}) =0$.  The transverse component of Eq.~(\ref{Pipi-3b}) (see the decomposition of Eqs.~(\ref{resultreal})) gives
\begin{equation}
\Pi_{b,(3)}^{T}(l_{0},{\bf l})  =  \frac{e^{2}}{1440 \pi^{2}} \left[-52\,{\bf l}^{2}+70l_{0}^{2}+30\frac{l_{0}^{4}}{{\bf l}^{2}}-15\frac{l_{0}^{3}}{{|\bf l|}^{3}}\left(l_{0}^{2}+2{\bf l}^{2}-3\frac{{\bf l}^{4}}{l_{0}^{2}}\right)  {\rm ln\,} \left({\frac{l_0+
|{\bf l}|}{l_0-|{\bf l}|}} \right) \right]
\end{equation}
while
\begin{equation}
\frac{l_i l_j}{{\bf l}^{2}} \Pi_{b,(3)}^{ij}(l_{0},{\bf l}) = \frac{l_0^2 }{|{\bf l}|^2} \,\Pi_{b,(3)}^{00}(l_{0},{\bf l}) + \frac{e^2}{2 \pi^2} \frac{{\bf l}^2}{30} \ ,
\end{equation}
from which we easily obtain Eq.(\ref{wid3}). 
The tadpole contribution  at order $1/p^3$ is then required to get the Ward identity fulfilled. 

To the particle contribution one should add the antiparticle contribution to the bubble diagram. One can show that at this order
\begin{equation}
{\widetilde \Pi}_{b,(3)}^{\mu\nu}(l) =\Pi_{b,(3)}^{\mu\nu}(l) \ .
\end{equation}

\subsubsection{Tadpole diagrams}

In this section, we show how two apparently equivalent ways to calculate the tadpole diagrams lead to different results.

\paragraph{Naive evaluation}\label{naive} $\quad$

If we proceed as in the previous sections,
the contribution of the tadpole diagrams reads

\begin{eqnarray}
\Pi^{\mu\nu}_{\rm t ,(3)}(l) &= & \frac{i}{2} \sum_{p, {\bf v}}\int\frac{d^{4} k }{(2\pi)^{4}} {\rm Tr}[W_{(3)}^{\mu \nu }(k,l)\, S_S^{(0)}(k) + W_{(2)}^{\mu \nu }(k,l)\, S_S^{(1)}(k) + W_{(1)}^{\mu \nu }\, S_S^{(2)}(k)] \, .
\label{tp}
\end{eqnarray}
Only the first term gives a dependence on $l$. 
Let us evaluate it in the following.
By substituting the zero-th order symmetric propagator in Eq.~(\ref{tp}) we obtain
\begin{equation}
\label{tadpole-3}
\Pi_{t, (3)}^{\mu\nu}(l)=  \frac {e^{2}}{2}\sum_{p, {\bf v}}\int \frac{d^{3}{\bf k}}{(2\pi)^{3}}
 {\rm Tr}( P_v \gamma_0   W_{(3)}^{\mu\nu}(k,l)) 
 (1-2  n_f(p) )
   \  .
\end{equation} 
After expressing this integral in terms of the momentum ${\bf q}$ according to Eq.~(\ref{def}), it is not
difficult to check that the pure thermal contribution is infrared (IR) divergent. In addition the $T=0$ contribution is both IR and UV divergent. However, 
the combination that appears in Eq. (\ref{tadpole-3}) is IR finite, as it can be seen by expanding for small $p$ the ratio $(1 - 2n_f(p))/p^3$.  
So the tadpole contribution at order $1/p^3$ is logarithmically divergent in the UV, but IR finite. The UV divergent piece fulfills the Ward identity, and hence it may be canceled by adding a proper counterterm in the 
Lagrangian built out of the different components of the electromagnetic field strength tensor.  
Furthermore,
finite contributions are also found, that added to 
 Eq.~(\ref{Pipi-3b}) result into a  polarization tensor which is respectful with the Ward identity.

 Let us see how this effectively works. We will use DR,  
with  $d= 3 + \epsilon$. We also neglect pieces that cancel after angular integration, so that the different tensorial components of Eq.(\ref{tadpole-3}) are

 \begin{eqnarray}
\Pi_{t, (3) }^{00} (l)& = & -\frac{e^{2} \mu^{3-d}}{4}\int  \frac{d^{d}{\bf q}}{(2\pi)^{d}}
\frac{1 -2 n_{f}(q)}{q^{3}}\,{\bf l}_{\perp}^{2}  \ ,
 \\
\Pi_{ t, (3)  }^{0i} (l)  & = & - \frac{e^{2}\mu^{3-d}}{4}\int  \frac{d^{d}{\bf q}}{(2\pi)^{d}}  \,\frac{1- 2n_{f}(q) }{q^{3}}\, l_{0}l_{\perp}^{i} \ ,
\\
\Pi_{t, (3)}^{ij} (l) & = & \frac{e^{2}\mu^{3-d}}{4}\int \frac{d^{d}{\bf q}}{(2\pi)^{d}}  \,\frac{1- 2n_{f}(q) }{q^{3}}\, 
\left ( \left(l_{0}^{2}+l_{\parallel}^{2}-{\bf l}_{\perp}^{2}\right)P^{ij}_T -l_{\parallel}\left(l_{\perp}^{i}v^{j}+l_{\perp}^{j}v^{i}\right) 
 -   {\bf l}_{\perp}^{2}v^{i}v^{j}\right )  \ .
\nonumber \\
 & &
\end{eqnarray}
 And after evaluation these give

\begin{eqnarray}
\Pi_{t, (3) }^{00} (l)& = &
\frac{e^{2}}{2\pi^{2}}\,\frac{1}{4}{\bf l}^{2}\left(\frac{2}{3\epsilon}+\frac{2}{3}\, \left( {\rm ln}\frac{\sqrt{\pi} T}{2 {\mu}}\, -\frac{\gamma}{2}-1 \right) +\frac{1}{9}\right) + 
{\cal O}(\epsilon)\,,\label{t003}\\
\nonumber \\
\Pi_{ t, (3)  }^{0i} (l)  & = & \frac{e^{2}}{2\pi^{2}}\,\frac{1}{4}l_{0}l^{i}\left(\frac{2}{3\epsilon}+\frac{2}{3}\,  \left( {\rm ln}\frac{\sqrt{\pi} T}{2 {\mu}}\, -\frac{\gamma}{2}-1 \right) 
+\frac{1}{9}\right) + {\cal O}(\epsilon)
\ ,
\\
\nonumber 
\end{eqnarray}
where ${\mu}$ is the renormalization scale, and $\gamma$ is Euler's constant.
The longitudinal and transverse components read

\begin{eqnarray}
 \frac{l_{i}l_{j}}{{\bf l}^{2}}\Pi^{ij}_{ t, (3)}  &=&
 \frac{e^{2}}{2\pi^{2}}\,\frac{1}{4}{ l_0^2}\left(\frac{2}{3\epsilon}+\frac{2}{3}\, 
 \left( {\rm ln}\frac{\sqrt{\pi} T}{2 {\mu}}\, -\frac{\gamma}{2}-1 \right) 
+\frac{1}{9}\right)  - \frac{e^{2}}{2\pi^{2}}\,\frac{{\bf l}^2}{30} +
{\cal O}(\epsilon)\,,
 \label{long}
\\
\frac {1}{2+\epsilon} \left(\delta_{ij}-\frac{l_{i}l_{j}}{{\bf l}^{2}}\right)\Pi^{ij}_{ t, (3)}  
&= & \frac{e^{2}}{2\pi^{2}}\,\frac{1}{4} { l_0^2}  \left(\frac{2}{3\epsilon}+\frac{2}{3}  \left( {\rm ln}\frac{\sqrt{\pi} T}{2 {\mu}}\, -\frac{\gamma}{2}-1 \right) 
\right) + \frac{e^{2}}{2\pi^{2}} \left( \frac{l_0^2}{36}
- \frac{{\bf l}^2}{15} \right)
\nonumber \\
&+ & {\cal O}(\epsilon)\ .
\label{trans}
\end{eqnarray}

The antiparticle contribution to the tadpole diagrams is found to be exactly the same as the particle contribution
\begin{equation}
{\widetilde \Pi}_{t,(3)}^{\mu\nu}(l) =\Pi_{t,(3)}^{\mu\nu}(l) \ .
\end{equation}

One can check now that the  the sum of the contributions of the bubble and tadpole diagrams obeys
\begin{equation}
l_\mu \left(\Pi^{\mu\nu}_{(3),b}  (l) + \Pi^{\mu\nu}_{(3),t} (l) \right)  =0 \ ,
\end{equation}
 and similarly, of course, for the antiparticle counterparts of these quantities.

The counterterms needed to remove the UV divergences differ from the QED vacuum ones (only the term proportional to $l_0^2$ in
Eq.~(\ref{long}) has the same UV divergence as in QED). We can write them as 
\be
{\cal L}_{c.t.}=-\frac{Z(\alpha , \epsilon) C(\alpha ,\mu)}{2}F_{0i}F^{0i}-\frac{Z'(\alpha , \epsilon) C'(\alpha ,\mu)}{4}F_{ij}F^{ij}
\label{c.t.} \ ,
\ee
where $Z$ and $Z'$ stand for the counterterms and $C$ and $C'$ stand for the matching coefficients.
\be
C=1+\frac{\alpha}{\pi}C^{(1)} \quad ,\quad C'= 1+\frac{\alpha}{\pi}C'^{(1)} \ .
\ee
From Eqs.~(\ref{t003})-(\ref{trans}) we need in the MS  renormalization scheme
\be 
Z=Z_{QED}=1-\frac{2}{3\epsilon}\frac{\alpha}{\pi}\quad ,\quad Z'= 1 \not= Z_{QED}\, ,
\ee
and, if we compare Eq.~(\ref{Pipi-3b}) and Eqs.(\ref{t003})-(\ref{trans}) with QED results of Appendix \ref{QED}, we see that they are identical if we choose
\be
C^{(1)}= 0 \quad ,\quad 
C'^{(1)}=\frac{2}{3}  \left( {\rm ln}\frac{\sqrt{\pi} T}{2 \mu}\, -\frac{\gamma}{2}-1 \right) 
+\frac{1}{9} \ .
\ee

Whereas there is nothing wrong in the fact that the UV counterterms of the effective theory differ from the ones of the fundamental one, it is indeed somewhat surprising in our case. At one loop, the calculation in the fundamental theory involves contributions from two particle (antiparticle) legs on-shell and from one particle (antiparticle) on-shell one antiparticle (particle) off-shell, as it has been made explicit in the Appendix \ref{QED}. There is a one-to-one mapping between these contributions and the bubble and tadpole diagrams of the EFT respectively.
Hence, at this order in $\alpha$, the EFT appears to be exactly equivalent to the fundamental theory and consequently one would expect the same UV behaviour. The results differ because the tadpole contribution is ambiguous even in dimensional regularization. The ambiguity becomes explicit if one,  for instance, puts $k-l$ rather than $k$ as the momentum in the loop. In the following section we devise a procedure by which the UV behavior of the fundamental theory is recovered while keeping the Ward identity fulfilled.

\paragraph{UV matched evaluation} \label{UV}$\quad$

\begin{figure}
\centering
\includegraphics{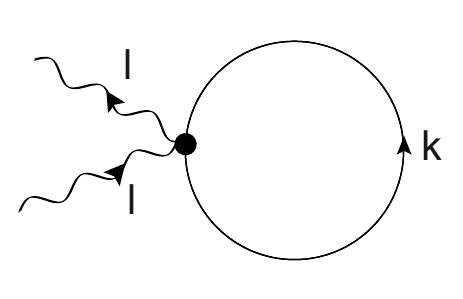}
\vspace{-0.5cm}\includegraphics{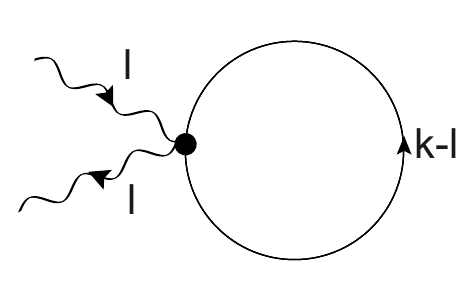}
\caption
{This figure illustrates how the tadpole diagrams are calculated in Sec. \ref{UV}.}
\label{tadfig}
\end{figure}


The rationale behind this procedure is that tadpole diagrams may be obtained from bubble diagrams in the fundamental theory by collapsing one of the legs. By doing so one obtains one tadpole with momentum $k$ in the loop and one tadpole with 
momentum $k-l$ in the loop, rather than only tadpoles with momentum $k$ in the loop, as we had in the section \ref{naive}. One prescription that provides tadpoles with momentum $k$ and $k-l$ in the loop is the following. When the incoming photon is to the left of the outgoing photon, we put $k$ as the momentum in the loop, and when it is the other way around, we put $k-l$ as the momentum in the loop, see Fig.~\ref{tadfig}.
Then formula (\ref{tp}) is replaced by

\begin{eqnarray}
\Pi^{\mu\nu}_{\rm t ,(3)}(l) &= &\Pi^{\mu\nu}_{\rm t ,(3),a}(l) +\Pi^{\mu\nu}_{\rm t ,(3,b)}(l) +\Pi^{\mu\nu}_{\rm t ,(3,c)}(l) \ ,
 \nn \\ 
\Pi^{\mu\nu}_{\rm t ,(3,a)}(l) &= & \frac{i}{2} \sum_{p, {\bf v}}\int\frac{d^{4} k }{(2\pi)^{4}} {\rm Tr}\left[W_{<(3)}^{\mu \nu }(k,l)\, S_S^{(0)}(k) + W_{>(3)}^{\mu \nu }(k-l,l)\, S_S^{(0)}(k-l) \right]\ ,
\nn\\
\Pi^{\mu\nu}_{\rm t ,(3,b)}(l) &= & \frac{i}{2} \sum_{p, {\bf v}}\int\frac{d^{4} k }{(2\pi)^{4}} {\rm Tr}\left[W_{<(2)}^{\mu \nu }(k,l)\, S_S^{(1)}(k) + W_{>(2)}^{\mu \nu }(k-l,l)\, S_S^{(1)}(k-l) \right] \ ,\\
\Pi^{\mu\nu}_{\rm t ,(3,c)}(l) &= & \frac{i}{2} \sum_{p, {\bf v}}\int\frac{d^{4} k }{(2\pi)^{4}} {\rm Tr}\left[ W_{<(1)}^{\mu \nu }\, S_S^{(2)}(k) + W_{>(1)}^{\mu \nu }\, S_S^{(2)}(k-l)\right] \ .
\nn
\label{tp2}
\end{eqnarray}
It turns out that only the pure spatial components are modified with respect to the naive prescription, so we will only provide the
explicit expressions for those below. 
\begin{eqnarray}
\Pi_{t, (3,a)}^{ij} (l)  &=&  \frac{e^{2}\mu^{3-d}}{4}\int \frac{d^{d}{\bf q}}{(2\pi)^{d}}  \,\frac{1- 2n_{f}(q) }{q^{3}}\, 
\Big ( \left(l_{0}^{2}+3 l_{\parallel}^{2}-2 {\bf l}_{\perp}^{2}\right)P^{ij}_T - 3l_{\parallel}\left(l_{\perp}^{i}v^{j}+l_{\perp}^{j}v^{i}\right) 
 +{\bf l}_{\perp}^{2}v^{i}v^{j} \Big) \ ,
  \nonumber\\
\Pi^{ij}_{\rm t,(3,b)}(l) &=& -  \frac{e^{2}}{2}\int \frac{d^{3}{\bf q}}{(2\pi)^{3}} \frac {1}{q^2}\frac{d n_f}{dq} 
\left(- 2 l^2_{\parallel}P^{ij}_T  +l _{\parallel}\left(l_{\perp}^{i}v^{j}+l_{\perp}^{j}v^{i}\right)  \right) \ , \\
\Pi^{ij}_{\rm t,(3,c)}(l) &=& -  \frac{e^{2}}{2}\int \frac{d^{3}{\bf q}}{(2\pi)^{3}} \frac{P^{ij}_T}{q} \left( \frac{1}{q}  \frac{d n_f}{dq} {\bf l}_{\perp}^{2} +  \frac{d^2 n_f}{dq^2} l^2_{\parallel}\   \right) \nn .
\end{eqnarray}
The two last equalities above are finite and need not be dimensionally regularized like the first one.
It turns out that the longitudinal component of the tadpoles 
is the same as the one obtained with the naive prescription Eq.~(\ref{long}).
However, the transverse part is modified 
so that Eq.~(\ref{trans}) becomes,
\begin{eqnarray}
&& \frac {1}{2+ \epsilon} \left(\delta_{ij}-\frac{l_{i}l_{j}}{{\bf l}^{2}}\right) \left(\Pi^{ij}_{\rm t, (3,a)} +  \Pi^{ij}_{\rm t, (3,b)} +\Pi^{ij}_{\rm t, (3,c)} \right) 
\nonumber
\\
 &=& \frac{e^{2}}{2\pi^{2}}\,\left(\frac{1}{\epsilon}+  \left( {\rm ln}\frac{\sqrt{\pi} T}{2 {\mu}}\, -\frac{\gamma}{2}-1 \right) 
\right)
\left(\frac{1}{6}(l_{0}^{2} - {\bf l}^2)\right) 
 + \frac{e^{2}}{2\pi^{2}} \left( \frac{l_0^2}{36} - \frac{{\bf l}^2}{90} \right) +
  {\cal O}(\epsilon) \ .
\end{eqnarray}

Antiparticles contribute in the exact same way as the particles, so as to compare with the full theory, we need to multiply by 2 the above result.

As advertised, we get now the same wave function renormalization as in QED, 
\be
Z=Z'=Z_{QED}=1-\frac{2}{3\epsilon}\frac{\alpha}{\pi} \ .
\ee
However, a non-vanishing matching coefficient at order $\alpha$ is still needed
to achieve agreement with the full theory result (see Appendix \ref{QED})
\be
C^{(1)}= 0 \quad ,\quad 
C'^{(1)}=\frac{1}{3} \ .
\ee

\subsubsection{Final result}

We display here the final results of our calculation for the polarization tensor, which upon the inclusion of the matching coefficients $C$ and $C'$, agree with the ones of the QED calculation of Appendix \ref{QED}. 
In the  MS renormalization scheme, and for 
 ${\mu}=\frac {\sqrt{\pi}}{2} T e^{-1 -\gamma/2}$, it reads

\begin{eqnarray}
\label{final-Long}
\Pi^{L}_{{\rm total},(3)} (l_{0},{\bf l}) &= &\frac{\alpha}{  \pi }\left[{\bf l}^{2}- \frac 13 l_{0}^{2}+ \frac 16 \frac{l_{0}}{|{\bf l}|}\left(l_{0}^{2}-3{\bf l}^{2}\right)  \left(
 \,{\rm ln\,}\left|{\frac{l_0+|{\bf l}|}{l_0-|{\bf l}|}}\right| 
-i \pi \, \Theta(|{\bf l}|^2 -l_0^2) \right)
 \right] \ , \\
 \Pi^{T}_{{\rm total},(3)} (l_{0},{\bf l}) &= & \frac{ \alpha}{\pi } \left[ \frac 12 l_{0}^{2} -\frac {2}{3} {\bf l}^{2}  + \frac 16\frac{l_{0}^{4}}{{\bf l}^{2}}- \frac {1}{12} \frac{l_{0}^{3}}{{|\bf l|}^{3}}
 \left(l_{0}^{2}+2{\bf l}^{2}-3\frac{{{\bf l}}^{4}}{l_{0}^{2}}\right)  \right.  \nonumber \\  
& \times& \left.   \left(
 \,{\rm ln\,}\left|{\frac{l_0+|{\bf l}|}{l_0-|{\bf l}|}}\right| 
-i \pi \, \Theta(|{\bf l}|^2 -l_0^2) \right)  \right] \ ,
\label{final}
\end{eqnarray}
where we have explicitly displayed the real and imaginary parts of the polarization tensors, the last corresponding to  corrections to HTL  Landau damping.
Let us comment here that our results of the longitudinal polarization tensor agree with the one-loop  computation of $\Pi^{00} ( 0, {\bf l})$ in Ref.~\cite{Blaizot:1995kg}
(see Eq.~(3.26)), see also Ref.~\cite{Andersen:1995ej}. To the best of our knowledge, the complete expression of the polarization tensor at this order in the $T$ expansion has not been computed before.

\section{Discussion}
\label{discussion}

We have shown how the EFT techniques that have been developed to study  different systems,
ranging from the high density regime  to the non-relativistic limits of QED and QCD, can also be applied to obtain power corrections to the HTL's at high temperature.
 We have used here the  OSEFT to systematically  organize the interactions of the hard scales of the plasma in powers of momenta,  a fact that  allows us
to recognize all the contributions to the one-loop diagrams to a given order in a $1/T$ expansion. Furthermore, with the OSEFT we can understand
the form of the non-localities that appear in these amplitudes at any order, as from the leading order Lagrangian Eq.~(\ref{Lan-1})   
we see that these can only be 
$1/iv.\partial$ or  $1/i{\tilde v}.\partial$ to a maximum power given by the order of the expansion (one at ${\cal O}(e^2T^2)$ and three
at ${\cal O}(e^2T^0)$, etc.).

Let us emphasize  that the OSEFT might have many other applications than those here presented. In particular, since it properly describes
the hard degrees of freedom of the plasma, it might be readily applied to the study of transport phenomena. Note also that all our  basic discussion of the derivation of 
the EFT Lagrangian in Sec. \ref{sec:OSEFT} does not require the presence of a thermal bath, and thus the OSEFT might have applications beyond thermal field theory. 

We should pin point here the differences and similarities that the OSEFT has with
respect to other  EFTs. In particular, the form of the OSEFT Lagrangian seems to be
quite similar to the Lagrangian of the HDET. The main difference relies on the fact that HDET is only strictly
valid at $T=0$, when there is a well-defined Fermi surface. The quantum fluctuations then are only those
around the Fermi surface. The high energy scale in HDET
is the chemical potential $\mu$, a fixed variable, and antiparticle fluctuations are not taken into account. In the OSEFT the high energy scale is the dynamical on-shell energy of the particles
or antiparticles, and these two degrees of freedom are treated on equal footing.
 HDET has been used to derive the so called hard dense loops in Ref.\cite{Schafer:2003jn},
  the explicit meaning of the sum given in the final expressions seems to differ
from the one 
in this paper. In that reference 
the sum is over the number of patches that cover 
the Fermi sphere, and 
an explicit
cutoff defining the maximal value of the residual momentum  
is introduced. Here, the sum is over hard momenta $p$ and the corresponding directions ${\bf v}$ and we can avoid
the introduction of an explicit cut-off by re-expressing all the final integrals in terms of the original full momentum variable.
 The OSEFT also shares many similarities with SCET, the main difference being that the latter is built for a fixed number of
privileged directions  along which the
particles are almost on shell (jet-like events), whereas in the OSEFT the almost on-shell particles may be found in any direction.

In this manuscript we have presented the first power correction in the high temperature expansion to the HTL polarization tensor in QED. 
As we already saw the contributions to the polarization 
tensor at
order $T$ vanish, and the first non-vanishing correction does not depend on $T$ (up to logarithms that fix the scale of the running coupling constant), even if it is due to the thermal effects in the plasma.
The new correction represents modifications of order $\alpha$, the electromagnetic fine structure constant, to the soft photon propagation.
This should be compared to the contributions to the photon polarization tensor arising at two-loop order from the hard scales, which are of order $e^4 T^2$.
Then for soft momenta, when  $l \sim eT$, the new contribution computed in this manuscript and the
two-loop order result would be equally important.

Our results can be readily applied to the computation of the electromagnetic polarization tensor in the quark-gluon plasma, by just taking into account the electromagnetic charges
of the different quark flavors. Again, at the QCD soft scale $l\sim gT$, where $g$ is the QCD gauge coupling constant, assumed to be small, the new contributions computed here would be of the same order as the  two-loop hard contribution, and hence a leading correction to the HTL result.

It might be worth it to compute the power corrections to the gluon polarization tensor in QCD.   The quark contribution to the gluon polarization tensor could be rescued from our QED result,
simply by taking into account some color  and flavor factors. The gluon contribution could be computed using similar ideas to those
here presented, although the proper framework to treat the gluons within the EFT 
should be first developed. In  QCD this would represent a next to leading order correction to the 
 the HTL polarization tensor (recall that in the case of QCD the soft contribution is Bose enhanced with respect to the hard one).
 
\acknowledgements

We thank Rob Pisarski and Juan Torres-Rincon, for a critical reading of our manuscript. We thank Stefano Carignano for detecting some numerical mistakes in an earlier
version of this manuscript.
We have been supported by the CPAN  CSD2007-00042 Consolider--Ingenio 2010 program, and the FPA2010-16963 and FPA2013-43425-P
projects (Spain). J.S. also acknowledges support from the 2014-SGR-104 grant (Catalonia) and the FPA2013-4657 project (Spain). He has also benefited from the INT-15-2c program {\it
Equilibration Mechanisms in Weakly and Strongly Coupled Quantum Field Theory}.
S.S. has been partially supported by the Schroedinger Fellowship of the FWF, project no. J3639 (Austria).


\appendix 

\section{Energy integration in the bubble-like diagrams}
\label{System-app}

In this Appendix we present the result of the expansion in large $p$ of the integral in  Eq.~(\ref{eq:IntExp}) after using the fermion dispersion law at second order, see 
Eq.~(\ref{displaw-2}). While $I_{k_{0}}^{(0)} =0$, we find
\begin{eqnarray}
i I_{k_{0}}^{(1)} & = & \frac{l_{\parallel}}{v\cdot l}\,\frac{dn_f}{dp},
\label{Ik-1}
\\
\nonumber \\
i I_{k_{0}}^{(2)} & = & \frac{1}{2p} \frac{dn_f}{dp}\,\left[-\frac{1}{\left(v\cdot l\right)^{2}}l_{\parallel}\left({\bf l}_{\perp}^{2}-2{\bf k}_{\perp}\cdot{\bf l}_{\perp}\right)- \frac{1}{v\cdot l}
\left({\bf l}_{\perp}^{2}-2{\bf k}_{\perp}\cdot{\bf l}_{\perp}\right)
\right] 
 \\
&+&  \frac{1}{2} \frac{d ^2n_f}{dp^2} \,\frac{1}{v\cdot l}\left(k_{\parallel}^{2}-(k_{\parallel}-l_{\parallel})^{2}\right)\,,
\nonumber \\
\nonumber 
\label{Ik-2}
\\
\label{Ik-3}
i I_{k_{0}}^{(3)} & = & \frac{1}{2p^{2}}  \frac{dn_f}{dp}\left[-\frac{1}{v\cdot l}\left(k_{\parallel}{\bf k}_{\perp}^{2}-(k_{\parallel}-l_{\parallel})({\bf k}_{\perp}-{\bf l}_{\perp})^2\right)+\frac{1}{2}\frac{1}{\left(v\cdot l\right)^{3}}l_{\parallel}({\bf l}_{\perp}^{2}-2{\bf k}_{\perp}\cdot{\bf l}_{\perp})^{2} \right.
\\
 & + &  \left.\frac{1}{2}\frac{1}{\left(v\cdot l\right)^{2}}\left(({\bf l}_{\perp}^{2}-2{\bf k}_{\perp}\cdot{\bf l}_{\perp})^{2}+ 2 l_{\parallel} \left((k_{\parallel}-l_{\parallel})({\bf k}_{\perp}-{\bf l}_{\perp})^2-k_{\parallel}{\bf k}_{\perp}^{2}\right) \right) \right]
 \nonumber \\
 & + & \frac{1}{2p} \frac{d^2n_f}{dp^2}\left[-\frac{1}{v\cdot l}\left((k_{\parallel}-l_{\parallel})({\bf k}_{\perp}-{\bf l}_{\perp})^2-k_{\parallel}{\bf k}_{\perp}^{2}\right)+\frac{1}{2}\frac{1}{\left(v\cdot l\right)^{2}}({\bf l}_{\perp}^{2}-2{\bf k}_{\perp}\cdot{\bf l}_{\perp})(l_{\parallel}^{2}-2k_{\parallel}l_{\parallel})\right]\nonumber \\
 & + & \frac{1}{6} \frac{d^3n_f}{dp^3}\frac{1}{v\cdot l}\left(k_{\parallel}^{3}-(k_{\parallel}-l_{\parallel})^{3}\right) \ ,
 \nonumber 
\end{eqnarray}
 and for retarded boundary conditions $l_0 \rightarrow l_0 + i \epsilon$.

The same quantities defined for the antiparticles, what we call $i {\tilde I}^{(n)}_{k_0}$, can be deduced from the particle's counterpart, applying the
basic rule of replacing $ p \rightarrow -p$, and also $\frac{d}{dp} \rightarrow - \frac{d}{dp}$, and ${\bf v} \rightarrow - {\bf v}$.

\section{OSEFT computations using the Imaginary Time Formalism}
\label{SecIFT}

The computations we carried out in this manuscript using the RTF can also be
reproduced using the ITF. 
In this Appendix we briefly mention  the main ingredients that
are needed to compute the OSEFT Feynman diagrams  in the ITF.

In order to proceed with the ITF one has to
 perform a rotation to Euclidean space-time of the theory.  One can derive the Euclidean propagators
 at every order in the $1/p$ expansion from our Euclidean rotated Lagrangians, where now the energies are  given by the
  fermionic Matsubara frequencies, $\omega_j = (2j +1) \pi T$, with $j \in Z$. It is also important to realize that the energy $p$ acts as a chemical potential 
  for the fermionic quantum fluctuations
  (or minus chemical potential for antifermionic fluctuations), see 
  Eq.~(\ref{eq:HamiltonianMu-1}), so that the Matsubara frequencies should be shifted accordingly in the propagators.

A major simplification of the computations using the ITF is also achieved if one performs the local field redefinitions
of Sec.~\ref{sec:Lag}.  Then the computation of  the different one-loop diagrams at a given order
 in $1/p$  basically involves the evaluation of two sorts of sums of Matsubara frequencies:  those that appear in the bubble diagrams, and those that appear
in the tadpole diagrams.
More particularly, for the bubble  diagrams there is always a sum over Matsubara frequencies of the form

\begin{equation}
T\sum_{j}\frac{1}{i\omega_{j}- p-f ({\bf k})}\frac{1}{i\omega_{j }- i \omega_s - p- f ({\bf k-l})} = - \frac{n_{f}(p+f({\bf k-l}))-n_{f}(p+f({\bf k}))}
{-i \omega_s - f ({\bf k-l}) + f ({\bf k})} \ ,\label{eq:ITF1}
\end{equation}
where $i\omega_s$ is the bosonic Matsubara frequency corresponding to the photon.
 Note that if we rotate back to Minkowski space $-i\omega_s \rightarrow l_0 + i\epsilon$, we recover the result of the basic integral
Eq.~(\ref{eq:IntExp}) which appears in the bubble diagrams using the RTF.

For the tadpole diagrams the only sort of Matsubara frequency sum to be considered is
\begin{equation}
T\sum_{j}\frac{1}{i\omega_{j}- p-f ({\bf k})}  = 1  - 2 n_{f}(p+f ({\bf k})) 
 \ ,\label{eq:ITF2}
\end{equation}
which also allows us to recover the same results for the tadpole diagrams computed with the RTF.

\section{Cancellation of the ${\bf k}^2$ terms}
\label{Cancellk2}

Here we consider  only the cancellation of the pieces of order ${\bf k}^2$ in the particle contributions to the tadpoles at order $1/p^3$, the same reasoning applies
to the antiparticle's contribution. As these pieces are the same no matter if one computes the tadpoles using the naive prescription, or the UV matched evaluation,
the proof applies to these two ways of computing the tadpoles. We concentrate on the tensorial structures which are spatial.

 Let us consider only the pieces which depend on ${\bf k}^2$ that appear in the computation at order $1/p^3$ in the tadpole diagrams. These read

\begin{eqnarray}
\Pi^{ij }_{\rm t ,(3,a)}(0) &= &\frac {e^{2}}{2}\sum_{p, {\bf v}}\int \frac{d^{3}{\bf k}}{(2\pi)^{3}}
\frac{  (1-2  n_f(p ) )}{p^3} \left (- P^{ij}_\perp ({\bf k}^2_\perp - 2 k^2_\parallel) + 2 k^i_\perp k^j_\perp - 2 v^i v^j {\bf k}^2_\perp 
\right.
\nonumber
\\
& -& \left. 4 k_\parallel 
(k^i_\perp v^j + k^j_\perp v^i ) \right ) \ ,
\label{annoyingk^2-A} \\
\label{annoyingk^2-B}
\Pi^{ij }_{\rm t ,(3,b)}(0) &= & 2 e^{2} \sum_{p, {\bf v}} \int \frac{d^{3}{\bf k}}{(2\pi)^{3}} \frac{1}{p^2} \Big ( - P^{ij}_\perp  k^2_\parallel + k_\parallel 
(k^i_\perp v^j + k^j_\perp v^i ) \Big ) \frac{dn_f}{dp} \ , \\
\label{annoyingk^2-C}
\Pi^{ij }_{\rm t ,(3,c)}(0) &= &\ e^{2} \sum_{p, {\bf v}} \int \frac{d^{3}{\bf k}}{(2\pi)^{3}} \frac{P^{ij}_\perp }{p} \left (- \frac{dn_f}{dp} \frac{{\bf k}^2_\perp}{p} -  \frac{d^2n_f}{dp^2}   k^2_\parallel  \right ) \ ,
\end{eqnarray}
These tadpole contributions can be trivially expressed in terms of the original variable ${\bf q}$, as to leading order ${\bf v} \sim {\bf \hat q}$, $n_f(p) \sim n_f (q)$, etc.
 We will see that  all of them are cancelled by the contributions arising from the lower order tadpoles in the $1/p$ expansion, when expressed in terms of the original momentum variable.

Let us consider the particle contribution to the tadpole diagram at order $1/p$, and re-express it in terms of the original momentum, 
 keeping pieces up to order $1/q^3$. More explicitly, after using Eqs.~(\ref{p-back}), (\ref{v-back}) and (\ref{n-back}), this tadpole diagram gives
\begin{eqnarray}
\label{re-expresstad1}
&&\Pi^{ij }_{\rm t ,(1)} =- e^2 \int\frac{d^{3}{\bf q}}{(2\pi)^{3}} \frac{1}{q} \left( 1  + \frac{k_{\parallel,{\bf q}}}{q}  - \frac{{\bf k}_{\perp,{\bf q}}^2 - 2 k^2_{\parallel,{\bf q}}}{2 q^2} \right) \left( 1 - 2 n_f(q) - 2 \frac{d n_f}{dq} \left( -  k_{\parallel,{\bf q}}  + \frac{{\bf k}_{\perp,{\bf q}}^2 }{2 q} \right)  \right. \nonumber \\
& -&\left. \frac{d^2 n_f}{dq^2} k_{\parallel, {\bf q}}^2 \right)
  \left( (\delta^{ij} - {\bf \hat q}^i {\bf \hat q}^j ) +  (1+ \frac{k_{\parallel,{\bf q}} } {q} ) \frac{{\bf \hat q}^i{\bf k}_{\perp,{\bf q}}^j   +{\bf \hat q}^j {\bf k}^i_{\perp,{\bf q}}} {q}
+  \frac{ 
 ({\bf \hat q}^i  {\bf \hat q}^j  {\bf k}_{\perp,{\bf q}}^2 -    {\bf k}_{\perp,{\bf q}}^i  {\bf k}^j_{\perp,{\bf q}})} {q^2}
 \right)  \ .
\end{eqnarray}
The pieces of order $1/q$ give account of the particle contribution to the tadpole diagram already considered in Sec.~\ref{Orderp1}
The terms of order $1/q^2$ cancel after performing the angular integral. We are then left with pieces of order 
$1/q^3$.   
  
 Even if the tadpole at order $1/p^2$, Eq.~(\ref{tp-2}), gives a vanishing contribution at order $e^2 T$, it still leads - after being expressed in terms
 of the original variable ${\bf q}$ - to contributions at order $1/q^3$, which have also to be considered.
More particularly, 
  Eq.~(\ref{tp-2}) expressed in terms of the original variables reads
\begin{eqnarray}
\Pi^{ij }_{\rm t ,(2,a)}(0) &=& - e^2 \int\frac{d^{3}{\bf q}}{(2\pi)^{3}} \frac{1}{q^2} \left( 1  + \frac{ 2k_\parallel^{\bf q}}{q}  \right) \left( 1 - 2 n_f(q) + 2  k_\parallel^{\bf q} \frac{d n_f}{dq}  \right)
\Bigg  [-  \left( k_\parallel^{\bf q} - \frac{{\bf k}_{\perp,{\bf q}}^2 }{ q} \right) \Big(  (\delta^{ij} - {\bf \hat q}^i {\bf \hat q}^j ) 
  \nonumber \\
  \label{re-expresstad2a}
&+ &   \frac{{\bf \hat q}^i{\bf k}_{\perp,{\bf q}}^j   +{\bf \hat q}^j {\bf k}^i_{\perp,{\bf q}}} {q} \Big)
-  (1+ \frac{k_\parallel^{\bf q} } {q} )({\bf \hat q}^i{\bf k}_{\perp,{\bf q}}^j   +{\bf \hat q}^j {\bf k}^i_{\perp,{\bf q}})
- \frac{ 2 ({\bf \hat q}^i  {\bf \hat q}^j  {\bf k}_{\perp,{\bf q}}^2 -    {\bf k}_{\perp,{\bf q}}^i  {\bf k}^j_{\perp,{\bf q}})} {q} \Bigg ]  \ ,
 \\
\Pi^{ij }_{\rm t ,(2,b)}(0) &=& 2 e^2 \int\frac{d^{3}{\bf q}}{(2\pi)^{3}} \frac{1}{q} \left( 1  + \frac{ k_\parallel^{\bf q}}{q}  \right) \left(  
\frac{d n_f}{dq}  \left( k_\parallel^{\bf q} -  \frac{{\bf k}_{\perp,{\bf q}}^2 }{ q} \right) -  \frac{d^2n_f}{dp^2}   k^2_{\parallel, {\bf q}}   \right)
 \Bigg [ (\delta^{ij} - {\bf \hat q}^i {\bf \hat q}^j ) 
 \nonumber
 \\
 \label{re-expresstad2b}
 &+& \frac{{\bf \hat q}^i{\bf k}_{\perp,{\bf q}}^j   +{\bf \hat q}^j {\bf k}^i_{\perp,{\bf q}}} {q} \Bigg ]\ , 
\end{eqnarray}
 which correspond to the first and second term of Eq.~(\ref{tp-2}), respectively. As mentioned in Sec.~\ref{Orderp2}  , the pieces of order $1/q^2$ above cancel after performing
 the angular integral.
 
 It is now easy to see that the sum of all the tadpole contributions at order $1/q^3$, Eqs.~(\ref{annoyingk^2-A}) to
 (\ref{re-expresstad2b}),  leads to a cancellation of the ${\bf k}^2$ dependence  at this order.

Similar computations should be carried out to see that the ${\bf k}^2$ pieces  in the bubble diagrams which appear at order $1/p^3$ cancel when re-expressing
the bubble contribution computed at lower orders in the $1/p$ expansion in terms of ${\bf q}$.

\section{The retarded polarization tensor in QED}
\label{QED}

In this Appendix we present the computation of the retarded polarization tensor in QED for soft external momentum $\sim eT$, and at the
same order of accuracy that was computed in this paper. We also use the RTF, and analyze and compare the result with that obtained with the
OSEFT. Let us recall that to leading order in a $T$ expansion one obtains the HTL, and that follows upon expanding
  the value of integrand of  the polarization tensor for
large values of the loop momentum, which is assumed to be of order $T$. 
 Subleading terms in the $T$ expansion of the polarization tensor can as well be obtained if one 
keeps subleading terms in the expansion of the integrand. This is the computation we have carried out to verify
the validity of our OSEFT results, and that we briefly summarize here.

In QED the retarded  photon polarization
tensor in the RTF reads  \cite{Carrington:1997sq} 
\begin{eqnarray}
\Pi^{\mu\nu}(l) &=&-\frac{ie^{2}}{2}\int\frac{d^{4}q}{(2\pi)^{4}}\Big ( {\rm Tr}[\gamma^{\mu}S_{S}(q^{\prime})\gamma^{\nu}S_{R}(q)]+{\rm Tr}[\gamma^{\mu}S_{A}(q^{\prime})\gamma^{\nu}S_{S}(q)]\Big) \nonumber \\
&- &\frac{ie^{2}}{2}\int\frac{d^{4}q}{(2\pi)^{4}}\Big ( {\rm Tr}[\gamma^{\mu}S_{A}(q^{\prime})\gamma^{\nu}S_{A}(q)]+{\rm Tr}[\gamma^{\mu}S_{R}(q^{\prime})\gamma^{\nu}S_{R}(q)]\Big) \ ,
\label{eq:IntDefFull} 
\end{eqnarray}
where  $q' = q- l$, and 
$S_{S}(q)$ and $S_{R/A}(q)$ are the 
electron propagators 
\begin{equation}
S_{R/A}(q)=\frac{\gamma\cdot q}{q^{2} \pm i\textrm{sgn}(q_{0})\epsilon} \, \qquad
 S_{S}(q)=-2\pi i\gamma\cdot q\,(1-2n_{f}(q_{0}))\delta(q^{2})\, ,\label{eq:propfull}
\end{equation}
and contain both the particle and antiparticle degrees of freedom.

The trace is easily evaluated 
\begin{equation}
J^{\mu \nu} [q;l] \equiv {\rm Tr}[\gamma^{\mu}\left(\gamma\cdot q^{\prime}\right)\gamma^{\nu}(\gamma\cdot q)]=4\left[q^{\mu}q^{\prime\nu}+q^{\prime\mu}q^{\nu}-g^{\mu\nu}q\cdot q^{\prime}\right]\,.\label{eq:TraceFull}
\end{equation}

The $q_0$-integration performed on the second line of Eq.~(\ref{eq:IntDefFull}) reduces to zero, as one can always close the countour in a half plane that does not contain a pole. We then consider the first term of  Eq.(\ref{eq:IntDefFull}) 

\begin{equation}
{\rm Tr}[\gamma^{\mu}S_{S}(q^{\prime})\gamma^{\nu}S_{R}(q)]=-8\pi i\left(1-2n_{f}(q_{0}^{\prime})\right)\delta\left(q^{\prime2}\right)\frac{1}{q^{2}+i\textrm{sign}(q_{0})\epsilon}
J^{\mu \nu} [q;l] \ .
\end{equation}
The denominator and delta function can be decomposed in the
following manner
\begin{eqnarray}
\frac{1}{q_{0}^{2}-\boldsymbol{q}^{2}+i\textrm{sign}(q_{0})\epsilon} & = & \frac{1}{2\left|\boldsymbol{q}\right|}\left(\frac{1}{q_{0}-\left|\boldsymbol{q}\right|+i\epsilon}-\frac{1}{q_{0}+\left|\boldsymbol{q}\right|+i\epsilon}\right)\,,\label{eq:PropDecomp}\\
\nonumber \\
\delta\left(q^{\prime2}\right) & = & \frac{1}{2\left|\boldsymbol{q}-\boldsymbol{l}\right|}\left\{ \delta\left[q_{0}-\left(l_{0}+\left|\boldsymbol{q}-\boldsymbol{l}\right|\right)\right]+\delta\left[q_{0}-\left(l_{0}-\left|\boldsymbol{q}-\boldsymbol{l}\right|\right)\right]\right\} \ .\nonumber \label{eq:DeltaDecomp}
\end{eqnarray}
These decompositions allows us to clearly identify the particle-particle, antiparticle-antiparticle and mixed contributions to the polarization tensor, a step that  help us
in our comparison with the OSEFT.  
We then arrive at 

\noindent 
\begin{eqnarray}
&&{\rm Tr}[\gamma^{\mu}S_{S}(q^{\prime})\gamma^{\nu}S_{R}(q)] =
-2i\pi \frac{\left(1-2n_{f}(\left|\boldsymbol{q}-\boldsymbol{l}\right|)\right)}{\left|\boldsymbol{q}\right|\left|\boldsymbol{q}-\boldsymbol{l}\right|} \times \\
&& \Bigg (
\left[\frac{1}{l_{0}+\left|\boldsymbol{q}-\boldsymbol{l}\right|-\left|\boldsymbol{q}\right|+i\epsilon}-\frac{1}{l_{0}+\left|\boldsymbol{q}-\boldsymbol{l}\right|+\left|\boldsymbol{q}\right|+i\epsilon}\right]J^{\mu \nu} \Big|_{q_{0}=l_{0}+\left|\boldsymbol{q}-\boldsymbol{l}\right|}  
 \label{eq:struc1}
 \nonumber
 \\
&& +  \left[\frac{1}{l_{0}-\left|\boldsymbol{q}-\boldsymbol{l}\right|-\left|\boldsymbol{q}\right|-i\epsilon}-\frac{1}{l_{0}-\left|\boldsymbol{q}-\boldsymbol{l}\right|+\left|\boldsymbol{q}\right|-i\epsilon}\right]J^{\mu \nu} \Big|_{q_{0}=l_{0}-\left|\boldsymbol{q}-\boldsymbol{l}\right|} \Bigg )\nonumber
\end{eqnarray}

Observe that each component of $J^{\mu \nu}$ 
depends on $q_{0}$ and is therefore modified by the delta function of the symmetric propagator in a different way, according to whether one considers the contribution of an on-shell
particle or antiparticle. A similar calculation has to be performed for the second term of
Eq.~(\ref{eq:IntDefFull}), which gives the contribution of on-shell particles and antiparticles carrying momentum ${\bf q}$ rather than ${\bf q}-{\bf l}$ as above. The fact that both ${\bf q}$ and ${\bf q}-{\bf l}$ 
on-shell momenta appear in the QED 
calculation suggests the prescription used in Sec. \ref{UV} to compute the tadpole diagrams in the OSEFT. 
Then one expands the resulting expressions for large $|{\bf q}|$. At leading order (${\cal O}(e^2T^2)$),  one obtains the HTL result. At 
${\cal O}(e^2T^1)$ the expressions can still be handled analytically, and lead to
the same result as provided by the OSEFT in Eq.~(\ref{Pibubble-2}), which  
cancels after performing the angular integral. At
${\cal O}(e^2T^0)$, there are
a large number of terms in the expansion, and we have carried out such a computation with the aid of a computer algebra system (Mathematica).

While at the lowest orders in the computation the structure of the bubble and tadpole diagrams that we encounter in the OSEFT is clearly seen,
at order $e^2 T^0$ the comparison with the OSEFT computation is not so straightforward.  In order
to reproduce the OSEFT structure of terms (that is, the same sort of integrals that appear in both the bubble and tadpole diagrams)
within   QED,  angular integrations have to be carried out, and also one has to integrate by parts the Fermi distribution function.
This applies to all orders in the $T$ expansion, but at this order things are more subtle.
This is in part due to the local field redefinitions we performed at order $1/p^2$ in the
OSEFT to simplify  the computations, which are clearly manifested at this order, and also to the appearance of logarithmic  UV divergences. 
For instance, if we call ``tadpole" in QED  those pieces whose integrand is proportional to $(1- 2n_f(q))$,  we see that there are also contributions arising from particle-particle and antiparticle-antiparticle interactions, and not only from mixed particle-antiparticle terms as it happens in the OSEFT. Let us call ``bubble" to the remaining contributions, which upon partial integrations become proportional to $d n_f(q)/dq$, we then have
\be
\Pi^{\mu\nu}(l)=
\Pi_{t'}^{\mu\nu}(l)+\Pi_{b'}^{\mu\nu}(l) \ .
\ee
The spatial components of the ``tadpole" contribution read
\ba
\Pi_{t'}^{ij}(l)&=&- e^2\mu^{3-d}\int\frac{d^{d}\boldsymbol{q}}{(2\pi)^{d}}\left\{ \frac{\left(1-2n_{f}\right)}{2|{\bf q}|^{3}}\,\left[3l_{\parallel}\left(l^{i}v^{j}+v^{i}l^{j}\right)-\left(l_{0}^{2}+7l_{\parallel}^{2}-3l_{\perp}^{2}\right)v^{i}v^{j}+\nonumber
\right. \right.\\ && \left.\left. +\,
\left(l_{0}^{2}+l_{\parallel}^{2}-2l_{\perp}^{2}\right)\delta^{ij}\right]\right\} \,.\label{eq:IJTP}
\ea
As in the OSEFT this expression is UV divergent, and it is regularized using DR, providing the photon wave function renormalization, as well as other finite contributions. The UV divergent terms agree with the ones obtained in Sec. \ref{UV} but disagree with the ones of Sec. \ref{naive}, as remarked before.  After regularization part of
the finite contributions can then also be expressed as a contribution proportional to the derivative of the Fermi distribution function, and hence  of the  form of the local pieces that may arise in the bubble contribution. 
The spatial components of the ``bubble" contribution in QED read
\begin{eqnarray}
\Pi^{ij}_{b'}(l) & = & -\frac{e^2}{2}\int\frac{d^{3}\boldsymbol{q}}{(2\pi)^{3}}\,\frac{1}{q^2}\frac{d n_f}{dq}\left(\left[2l_{\parallel}+\left(l_{\perp}^{2}-3l_{\parallel}^{2}\right)\left(\frac{1}{v\cdot l}-\frac{1}{\tilde{v}\cdot l}\right)
\right.\right.\label{eq:IJNL}\\
 &+& \left.\left. \,
l_{\perp}^{2}l_{\parallel}\left(\frac{1}{\left(v\cdot l\right)^{2}}+\frac{1}{\left(\tilde{v}\cdot l\right)^{2}}\right)\right]\left(l^{i}v^{j}+v^{i}l^{j}\right)+\,\left[2l_{\perp}^{2}-8l_{\parallel}^{2}-\left(8l_{\perp}^{2}l_{\parallel}-\frac{22}{3}l_{\parallel}^{3}\right)
\right.\right.\nonumber\\
&\times & \left.\left.
\left(\frac{1}{v\cdot l}-\frac{1}{\tilde{v}\cdot l}\right)
 +\,
\left(l_{\perp}^{4}-5l_{\parallel}^{2}l_{\perp}^{2}\right)\left(\frac{1}{\left(v\cdot l\right)^{2}}+\frac{1}{\left(\tilde{v}\cdot l\right)^{2}}\right) + l_{\perp}^{4}l_{\parallel}\left(\frac{1}{\left(v\cdot l\right)^{3}}-\frac{1}{\left(\tilde{v}\cdot l\right)^{3}}\right)\right]v^{i}v^{j} \nonumber 
\right.\nonumber\\ &+&\left. \,
 \left[4l_{\parallel}^{2}-2l_{\perp}^{2}+l_{\parallel}l_{\perp}^{2}\left(\frac{1}{v\cdot l}-\frac{1}{\tilde{v}\cdot l}\right)\right]\delta^{ij}\right) \ ,
 \nonumber 
\end{eqnarray}
 where the particle and antiparticle contributions are displayed. The latter, after performing the change of variables ${\bf v} \rightarrow - {\bf v}$ in the integral, can be expressed
in the same way as the particle contribution.
The non-local pieces of the above expression agree with  twice the non-local pieces of Eq.~(\ref{Pipi-3b}), whereas the local pieces above add up to zero upon angular integration.

For $\Pi^{00}(l)$ and $\Pi^{0i}(l)$ we obtain exactly the same expressions as in Sec. \ref{Orderp3} both for bubble and tadpole contribution, and also for UV divergent and finite pieces.

In summary, we have  checked that  the OSEFT reproduces the polarization tensor of QED, up to a local piece at order $e^2 T^0$. This requires the addition of the electric and magnetic terms of the Maxwell Lagrangian multiplied by suitable  matching coefficients, as discussed in Secs. \ref{naive} and \ref{UV}. The final result for the longitudinal and transverse components of the polarization tensor is displayed in  Eqs.~(\ref{final-Long})-(\ref{final}).

\end{document}